\documentclass[fleqn,usenatbib]{mnras}

\usepackage{newtxtext,newtxmath}

\usepackage[T1]{fontenc}
\usepackage{ae,aecompl}

\usepackage{graphicx}	
\usepackage{amsmath}	
\usepackage{bm}
\usepackage[usenames]{color}
\usepackage{animate}
\usepackage{mathrsfs}

\graphicspath{{Figures/}}
\usepackage[normalem]{ulem}
\usepackage{array}

\newcolumntype{P}[1]{>{\centering\arraybackslash}p{#1}}
\newcolumntype{M}[1]{>{\centering\arraybackslash}m{#1}}

\usepackage[normalem]{ulem}

\title[Stokes lensing]{Imaginary images and Stokes phenomena in the weak plasma lensing of coherent sources}

\author[Jow et al.]{
Dylan L. Jow,$^{1,2}$\thanks{E-mail: djow@physics.utoronto.ca}
Fang Xi Lin,$^{1,2}$
Emily Tyhurst,$^{1,2}$
and Ue-Li Pen$^{1,2,3,4,5}$
\\
$^{1}$Canadian Institute for Theoretical Astrophysics, University of Toronto, 60 St. George Street, Toronto, ON M5S 3H8, Canada\\
$^{2}$Department of Physics, University of Toronto, 60 St. George Street, Toronto, ON M5S 1A7, Canada\\
$^{3}$Perimeter Institute for Theoretical Physics, 31 Caroline St. North, Waterloo, ON, Canada N2L 2Y5\\
$^{4}$Canadian Institute for Advanced Research, CIFAR program in Gravitation and Cosmology\\
$^{5}$Dunlap Institute for Astronomy \& Astrophysics, University of Toronto, AB 120-50 St. George Street, Toronto, ON M5S 3H4, Canada
}

\date{Accepted XXX. Received YYY; in original form ZZZ}

\pubyear{2021}

\begin{document}
\label{firstpage}
\pagerange{\pageref{firstpage}--\pageref{lastpage}}
\maketitle

\begin{abstract}
The study of astrophysical plasma lensing, such as in the case of extreme scattering events, has typically been conducted using the geometric limit of optics, neglecting wave effects. However, for the lensing of coherent sources such as pulsars and fast radio bursts (FRBs), wave effects can play an important role. Asymptotic methods, such as the so-called Eikonal limit, also known as the stationary phase approximation, have been used to include first-order wave effects; however, these methods are discontinuous at Stokes lines. Stokes lines are generic features of a variety of lens models, and are regions in parameter space where imaginary images begin to contribute to the overall intensity modulation of lensed sources. Using the mathematical framework of Picard-Lefschetz theory to compute diffraction integrals, we argue that these imaginary images contain as much information as their geometric counterparts, and may potentially be observable in data. Thus, weak-lensing events where these imaginary images are present can be as useful for inferring lens parameters as strong-lensing events in which multiple geometric images are present. 
\end{abstract}




\section{Introduction}
\label{sec:intro}

Since the first detections of extreme scattering events (ESEs) in which radio sources undergo large, frequency dependent changes in intensity, there have been significant ongoing efforts in the study of astrophysical plasma lensing. While the first observations of ESEs were of quasars \citep{Fiedler1987}, pulsars were also observed to undergo similar changes in intensity \citep{1993Natur.366..320C}. Following the first observations of these events which were classified under the broad heading of ``extreme scattering events", it was proposed that large-scale inhomogeneities in the interstellar medium (ISM) refracted the light from the sources, resulting in the observed intensity modulations \citep{1994ApJ...430..581F, 1996ApJ...457L..23C}. Since then, significant efforts to model these events and infer the properties of the lenses responsible for observations have been made \citep[see e.g.][]{1987Natur.328..324R,1998ApJ...496..253C, 2013A&A...555A..80P, 2016ApJ...817..176T, 2016Sci...351..354B, 2018Natur.557..522M}. A variety of different types of lens models have been proposed as generically responsible for ESEs, with observational consequences for these models being suggested \citep[see e.g.][]{2012MNRAS.421L.132P, 2018MNRAS.475..867E, 2018MNRAS.481.2685D, simard_predicting_2018}.

To date, much of the work on astrophysical plasma lensing has been carried out using the geometric optics formalism. Since most astrophysical sources are not coherent \citep{Schneider}, geometric optics is an adequate description of plasma lensing in a wide variety of cases. Moreover, the full wave optics description requires the computation of highly oscillatory diffraction integrals, which are often analytically and numerically challenging to perform. However, for sources such as pulsars and fast radio bursts (FRBs), which are effectively coherent point-sources in many practical contexts, wave effects will be important in describing their lensing behaviour. Recently, the number of observations of ESEs in pulsars has been growing \citep{Coles2015, 2018ApJ...861..132L}, and while ESEs have not been definitively observed for FRBs, plasma lensing has been suggested as being responsible for certain features in FRB data \citep{2017ApJ...842...35C}. 

Recently, some work has been done to take into account wave effects in the lensing of coherent sources. \citet{2018Natur.557..522M} use a simple elliptic lens model to model wave effects in the lensing of PSR B1957+20 by its companion, and \citet{2019MNRAS.484.5723L} use wave optics to constrain properties of the local magnetic fields for the same pulsar. \citet{GrilloCordes2018} describe an asymptotic method to accurately compute wave effects near caustics (regions where the geometric optics limit results in diverging intensity modulations). One of the reasons for the ubiquity of geometric optics is the fact that it reduces the study of lensing to computing the intensities of a discrete set of isolated images. The mathematical framework of catastrophe theory allows one to describe the ways in which these images form, and how they behave near caustics (also known as \textit{catastrophes}). A powerful result from catastrophe theory is that these caustics must take on one of a few standard forms, which do not strongly depend on the precise details of the underlying lensing potential. This makes geometric optics an extremely powerful tool for understanding lensing, both computationally and conceptually \citep{Nye}. As \citet{GrilloCordes2018} point out, the advantage of using an asymptotic method for computing wave effects is that it preserves the geometric optics point of view of a discrete set of images, each with their own intensity modulations. Other methods of computing diffraction integrals that arise in wave optics do not generally make reference to images, and so lose some of the power of catastrophe theory in describing lensing. 

The asymptotic method used by \citet{GrilloCordes2018} is useful for computing wave effects near fold catastrophes, which generically occur in strong-lensing regimes, where multiple images undergo large intensity modulations. Much of the work on describing the geometric optics of ESEs have also focused on these multiple-image, strong-lensing regimes. When multiple images are present, intensity modulations are generally much larger, and so more likely to be seen in data. Moreover, more information can be inferred about a lens when multiple images are present. The goal of this paper will be to further the study of wave optics in weak-lensing regimes, i.e. where there is only one geometric image. In particular, we will argue that, even in weak-lensing regimes, wave effects may be observable, and can contain information that has previously been neglected. These single geometric image regimes are generic, as, in the case of plasma lensing, they will occur above a certain frequency for any lensing event.

To that end, we will draw upon the mathematical framework of Picard-Lefschetz theory, which provides an exact approach to computing highly oscillatory integrals, and was recently applied to diffraction integrals in lensing \citep{job_pl}. Unlike the asymptotic method described by \citet{GrilloCordes2018}, which is an approximate method that is valid only near catastrophes of a specific form, Picard-Lefschetz theory is exact and general. In brief, Picard-Lefschetz theory analyses diffraction integrals by extending the integrand into the complex plane in order to provide a numerically stable means of computing diffraction integrals for a large variety of lenses. Picard-Lefschetz theory also has the advantage of preserving the geometric picture of a discrete set of images, as the final contour in the complex plane can be broken up into a discrete set of sub-contours (called ``Lefschetz thimbles") which correspond to the geometric images. In addition to Lefschetz thimbles corresponding to geometric images, there will generically be contributions from imaginary images, which are neglected in geometric optics.

The goal of this paper is twofold: to demonstrate the utility of Picard-Lefschetz theory (not only as a numerical method, but also as a conceptual framework that bridges the gap between wave optics and geometric optics), and to demonstrate that in weak-lensing regimes imaginary images can significantly contribute to the total intensity modulation induced by a lens. We will argue that these imaginary images are generic and contain as much information about the lens as real images, and may, in principle, be observable in data. In Section~\ref{sec:wave}, we will briefly introduce the lensing formalism, comparing and contrasting the full wave optics description with geometric optics. We will also describe, in more detail, Picard-Lefschetz theory, and the role of imaginary images in wave optics. Though our primary motivation for studying these phenomena is plasma lensing, the formalism described is general and may be applied to gravitational lensing, as well. In Section~\ref{sec:rational}, we will study a specific lens model, the rational lens, in order to elucidate the concepts introduced in Section~\ref{sec:wave} with a concrete example. We will show that even in the weak-lensing regime for the rational lens, imaginary images can produce significant diffractive effects, which can break degeneracies between the underlying physical parameters of the lens. Finally, in Section~\ref{sec:prospects} we will describe the prospects for observing imaginary images in weak-lensing data.

\section{Wave optics in Lensing}
\label{sec:wave}



\begin{figure}
    \centering
    \includegraphics[width=\columnwidth, trim = 0 40 120 50]{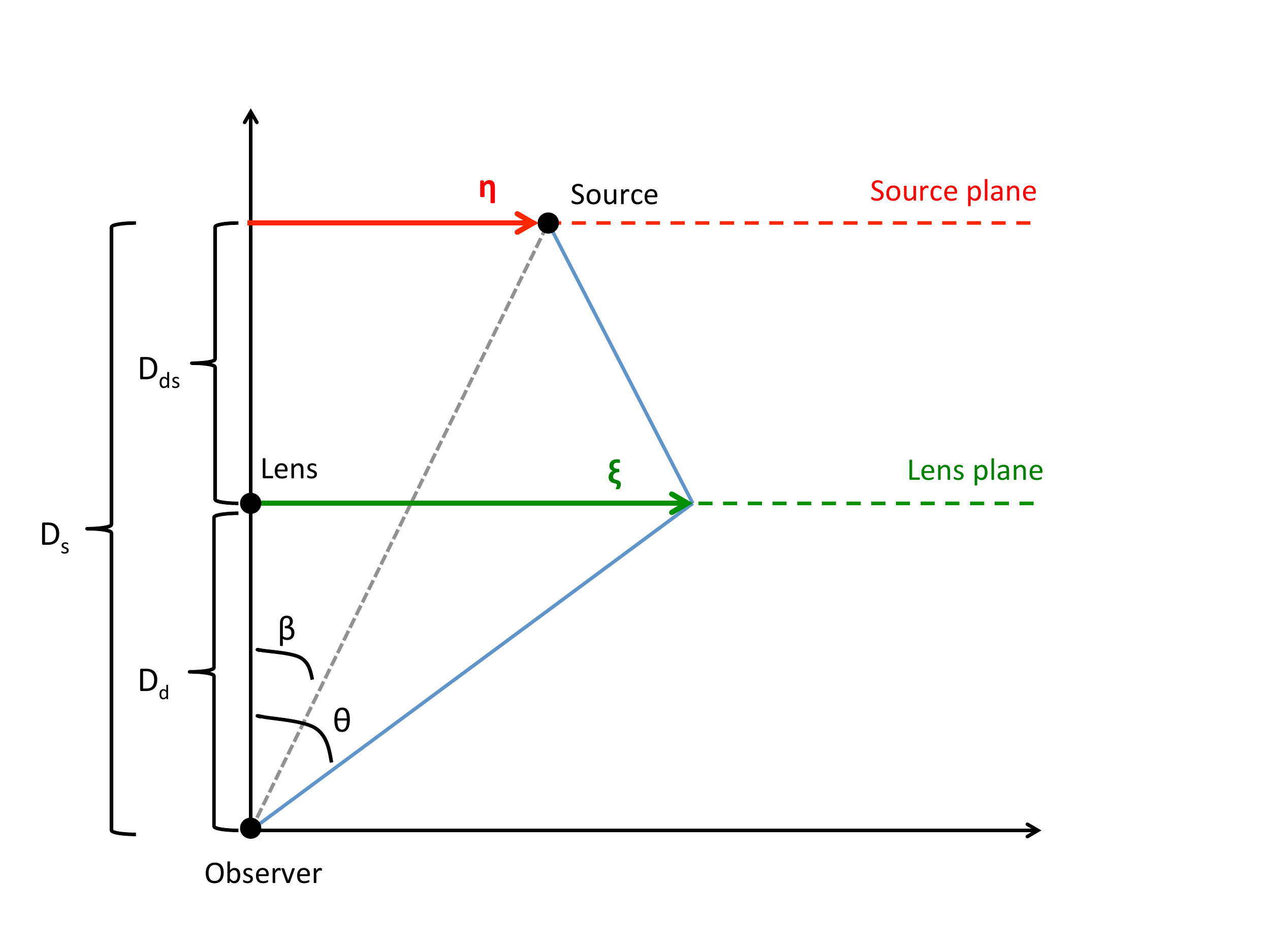}
    \caption{
    Geometry of a source at a distance $D_s$ from an observer being lensed in the thin-lens approximation by a lens at a distance $D_d$ from the observer. The unperturbed line-of-sight from observer to source is shown as a gray dashed line. The vectors $\bm{\eta}$ and $\bm{\xi}$ are vectors in the source and lens plane, respectively (planes perpendicular to the optical axis, containing the source and lens). Together, $\bm{\eta}$ and $\bm{\xi}$ define a path, shown in blue, from source to observer. The radiation seen by the observer is determined by a path integral over all such paths.
    }
    \label{fig:lens_setup}
\end{figure}

Lensing occurs when a differential phase difference is induced in a wave propagating through a medium. In the thin-lens approximation, the medium inducing this phase difference is restricted to a single plane at a fixed distance between the source and the observer. For a monochromatic wave emanating from a point source, we can write this phase difference as 
\begin{equation}
    S({\bm \theta}, {\bm \beta}) = \omega \Big[ \frac{D_d D_s}{2 c D_{ds}} |{\bm \theta} - {\bm \beta}|^2 - \psi ({\bm \theta}, {\bm \omega}) \Big],
    \label{eq:phase_diff}
\end{equation}
where $\omega$ is the angular frequency of the wave and the angles and distances involved are shown in the generic lensing diagram in Fig.~\ref{fig:lens_setup}. The first term is simply the phase difference induced by the geometric difference in length between different paths from source to observer compared to a straight line between source and observer (i.e. the difference in length between the blue line and the grey dashed line in Fig.~\ref{fig:lens_setup}). The second term is the phase difference induced by the lensing medium and is encapsulated by the lensing potential $\psi({\bm \theta}, \omega)$. For a gravitational lens, this potential is independent of $\omega$, and is due to the time delays induced by the curvature of space-time. For a plasma lens, $\psi({\bm \theta}, \omega) = \Sigma_e({\bm \theta}) e^2 / 2 m_e \epsilon_0 \omega^2$, where $\Sigma_e$ is the electron surface density in the lens plane. The optical axis used to define the angles ${\bm \theta}$ and ${\bm \beta}$ is arbitrary, but is typically chosen such that the potential $\psi$ has its maximum at ${\bm \theta} = 0$. The term inside the bracket of Eq.~\ref{eq:phase_diff} can be interpreted as a time delay between light traveling along two different paths through the lens.

When the source is a coherent, monochromatic point source, the field at the observer is given by
\begin{align}
\nonumber
    F(\omega, \bm{\beta}) &= \frac{\omega}{2\pi c i} \frac{D_dD_s}{D_{ds}} \int d^2\bm{\theta} e^{i S({\bm \theta}, {\bm \beta})} \\
    &= \frac{1}{2\pi i \theta^2_F} \int d^2\bm{\theta} e^{i S({\bm \theta}, {\bm \beta})},
\label{eq:F}
\end{align}
where the normalization is chosen such that $F = 1$ when $\psi = 0$. The Fresnel scale is defined as $\theta_F = \sqrt{c/\omega D}$, where $D = D_d D_s / D_{ds}$, and is critical in determining the scale of interference effects. The observed intensity of the field is then $|F|^2$. This result is derived in \citet{Schneider} in the context of gravitational lensing, as well as in \citet{Nye} in a more general context. \citet{job_pl} derives Eq.~\ref{eq:F} using a path integral formulation familiar from quantum mechanics. Eq.~\ref{eq:F} is for the lensing of monochromatic plane waves. Thus, in order to apply, the time delays must be much smaller than the coherence time of the source. 

Understanding lensing phenomena essentially boils down to computing the diffraction integral in Eq.~\ref{eq:F}. However, numerically computing highly oscillatory integrals of this form is not generally an easy task, and except for a few simple lensing potentials, few analytic solutions exist. Historically, the diffraction integral has not been computed at all; rather lensing phenomena have been studied in the geometric limit of optics. In the geometric limit, optical phenomena are studied by looking at the geometry of light rays (i.e. lines perpendicular to the phase surface of the waves). An observer sees a discrete set of images (rays connecting source to observer) and the intensity of each image is determined by the focusing or anti-focusing of rays around the image. The geometric limit does not take into account the wave nature of light, and so cannot account for interference effects. However, since coherence is a prerequisite for wave effects, and since most astrophysical sources are not coherent, the geometric limit is valid for a variety of applications, hence its ubiquity. 

The Eikonal limit adds first-order wave effects to the geometric limit. As in the geometric limit, an observer sees only a discrete set of images; however, the phase along the rays corresponding to each image are computed, and the images are allowed to interfere with each other at the observer. Mathematically, the Eikonal limit is equivalent to the stationary-phase approximation of Eq.~\ref{eq:F}, and is valid in the high frequency limit. In the Eikonal limit, only a finite set of paths - the images - contribute to the diffraction integral. For low frequencies, this approximation breaks down as the contribution from a non-finite set of paths becomes non-negligible. When this happens, it is not possible to assign the contributions to the flux to a discrete set of images, and Eq.~\ref{eq:F} must be computed in its entirety.

The concept of images, however, is an extremely powerful tool, as understanding when and how images are formed can allow for the classification of lenses according to catastrophe theory, that does not depend on the precise details of the lensing potential \citep{Nye}. This is in contrast to Eq.~\ref{eq:F}, which, on its own, is rather opaque and gives little insight into the underlying physics. Recently, Picard-Lefschetz theory has been introduced as a tool to compute these diffraction integrals \citep{job_pl}. In addition to its power as a numerical tool, the mathematics of Picard-Lefschetz theory can help bridge the conceptual gap between wave optics and geometric optics.

In the following sections, we will describe the geometric and Eikonal limits in more detail, as well as how Picard-Lefschetz theory can be applied numerically and conceptually to make computing and understanding Eq.~\ref{eq:F} in the full wave regime more tractable.

\subsection{Geometric optics and the Eikonal Limit}
\label{sec:geom}



In the high-frequency limit, the diffraction integral is dominated by a discrete set of isolated images. In this limit, the resultant intensity rapidly oscillates as a function of source position and frequency due to the interference between the images. Geometric optics is obtained when these oscillations are averaged over, which may physically occur due to the incoherence of the source. The geometric images are given by the rays that connect the source and the observer, which, using the geometry shown in Fig.~\ref{eq:F}, are given by \citep{Nye}
\begin{equation}
    \nabla_{\bm \theta} S({\bm \theta}, {\bm \beta}) = 0.
\end{equation}
In other words, the geometric images correspond to the points of stationary phase along the wave front. In gravitational lensing, where the group delay and phase delay are the same, this is equivalent to Fermat's principle of least time, in which a photon is taken to travel along the path that extremizes the time between source and observer.

For the sake of simplicity, we can rewrite the phase difference, Eq.~\ref{eq:phase_diff}, in terms of dimensionless quantities. Let $\theta_0$ be some characteristic scale of the lens, then write ${\bm x} = {\bm \theta}/\theta_0$, ${\bm y} = {\bm \beta}/\theta_0$, and $\nu = \theta_0^2 / 2 \theta_F^2$. Then Eq.~\ref{eq:phase_diff} can be written as
\begin{equation}
    S({\bm x}, {\bm y}; \nu) = \nu \Big[({\bm x} - {\bm y})^2 - \hat{\psi}({\bm x}; \nu) \Big],
    \label{eq:dimless_S}
\end{equation}
where $\hat{\psi}$ is the dimensionless lensing potential. The lens equation is then
\begin{equation}
    {\bm y} \equiv {\bm \xi}(\bm x) = {\bm x} - \frac{1}{2} \nabla_{\bm x} \hat{\psi}({\bm x}; \nu).
\end{equation}
The function ${\bm \xi}({\bm x})$ determines a mapping between positions in the lens plane, ${\bm x}$, and positions in the source plane, ${\bm y}$. For a given source position, ${\bm y}$, the geometric images correspond to the real solutions $\{ {\bm x}_i \}$ of the lens equation. 

Each image is magnified by an amount that can be determined by the geometry of the rays around each image. In particular, the magnification of image ${\bm x_i}$ is given by \citep{Schneider}
\begin{equation}
    |F_i|^2 = \Big| \det \frac{\partial {\bm \xi}_j({\bm x_i})}{\partial {\bm x}_k} \Big|^{-1} \equiv |\Delta_i |^{-1}.
\end{equation}
This can be seen intuitively, since the Jacobian $\frac{\partial {\bm y}}{\partial {\bm x}}$ determines the change in area an image of size $d{\bm x}$ in the lens plane undergoes when it is mapped to the source plane. This change in area determines how much the rays are focused or anti-focused, which determines the magnification induced by the lens. The total magnification a point source at position ${\bm y}$ undergoes is then
\begin{equation}
    |F|^2_\mathrm{geom.} = \sum_i |F_i|^2.
\end{equation}
The $|.|^2$ is to indicate that these are intensity modulations with no phase information. 

The geometric magnification is often straightforward to compute for a variety of lens models and ultimately boils down to finding solutions to the lens equation. A necessary condition for geometric optics to hold is that the source be non-coherent, which happens to be true for most astrophysical sources. Incoherence of the source physically results in the averaging over the oscillations in intensity as a function of source position and frequency, by which the geometric limit is obtained from the full diffraction integral \citep{Schneider}. Most sources are either incoherent due to the details of their emission, or they are rendered effectively incoherent due to their size. For example, sources whose angular size is on the order of, or larger than the Fresnel scale, $\theta_F$, do not produce interference fringes \citep{Jow2020}. 

Even for non-coherent sources, geometric optics breaks down at caustics, which are given by source positions ${\bm y}$ such that $\Delta_i = 0$ for one or more images. When this occurs, the magnification is formally infinite, which is unphysical. While physically there must always be some maximum allowed magnification, caustics correspond to regions of high magnification in observations. Since caustics correspond to singularities in the lens mapping, regions on the source plane where the lens equation has different numbers of real solutions are separated by caustics.

These caustics are the catastrophes of the lens mapping, and are amenable to study using the powerful mathematical tool of catastrophe theory. Catastrophe theory allows us to classify caustics according to a small number of elementary forms \citep{Nye}. Studying these canonical forms allows us to learn much about an optical system without needing to know the precise details of the lensing potential. 

The Eikonal limit is essentially geometric optics but taking into account first-order wave effects. The Eikonal limit only takes into account the contributions from a discrete set of images, but, in addition to computing the magnification of the individual images, one also computes the phase of each image at the observer. The result is that the observed field at an observer due to each individual image is \citep{nakamura}
\begin{equation}
    F_i({\bm y}; \nu) = \frac{1}{|\Delta_i|^{1/2}} e^{iS({\bm x}_i, {\bm y}; \nu) - i \frac{\pi}{2} n_i},
\end{equation}
where $n_i = -1, 0, 1$ when $x_i$ is a minimum, saddle point, or maximum of $S({\bm x}, {\bm y}; \nu)$. The total field is the sum of this, and the total magnification is given by
\begin{equation}
    |F|^2_\mathrm{Eik.} = \big| \sum_i F_i \big|^2.
\end{equation}
Thus, the Eikonal limit is simply the geometric limit, but the individual images are allowed to interfere. 

Mathematically, the Eikonal limit is equivalent to the stationary-phase approximation of the integral in Eq.~\ref{eq:F}. In the stationary-phase approximation, the oscillatory nature of the integrand means that in the limit $\nu \to \infty$, only the points $x_i$ for which the phase function is stationary contribute to the integral. The Eikonal limit is, therefore, a good approximation of the full wave regime in the high $\nu$ limit. However, just as geometric optics breaks down when $\Delta_i = 0$ for some image, so too does the Eikonal limit, leading to a divergent magnification. The Eikonal limit is not valid near caustics; however, as $\nu$ gets larger, the region around the caustics for which the Eikonal limit fails gets smaller. 

So far, in this brief overview of geometric optics and first-order wave optics, we have assumed that the only images that contribute to the field at the observer are the real solutions to the lens equation, Eq.~\ref{eq:dimless_S}, which we call the ``geometric images". However, the lens equation may have imaginary solutions, and, indeed, under certain conditions these imaginary images may play an important role. We will hold off on the discussion of imaginary images until Section~\ref{sec:imaginary_images}, and turn now to a numerical method for evaluating the full wave optics integral.

\subsection{Picard-Lefschetz Theory}
\label{sec:pl}

We wish to compute oscillatory integrals of the form
\begin{equation}
    I = \int_{\mathbb{R}^n} d^n {\bm x} e^{i S({\bm x}; {\bm \mu})},
\end{equation}
where ${\bm \mu}$ is a set of parameters that fix the phase function $S({\bm x})$. Such oscillatory integrals arise in a variety of areas in physics, diffraction integrals (Eq.~\ref{eq:F}) being one such area. Integrals of this type are conditionally convergent, and so a naive attempt at numerical integration will fail, as this would require a sampling of the entire (infinite) domain. For analytic phase functions whose real part is a Morse function, Picard-Lefschetz theory guarantees the existence of a surface in the complex plane $\mathbb{C}^n$ which is a continuous transformation of the original integration domain, $\mathbb{R}^n$, where the integrand is localized and non-oscillatory. 

To see how this works, we define the function $h({\bm x}) = \text{Re}\{i S({\bm x})\}$, which we take to be a Morse function. A Morse function is any real function on a smooth manifold with non-degenerate critical points (for an overview of Morse theory with applications to gravitational lensing, see \citet{2001stgl.book.....P}). For the remainder of this section, we will refer to the Morse function as $h({\bm z})$, to encode the analytic continuation of the phase function from the real domain, $\mathbb{R}^n$, to the complex plane, $\mathbb{C}^n$. A result of Morse theory is that if $h$ is a Morse function that is the real part of an analytic function, the critical points of $h$ all have Morse index $n$, which is the number of independent directions away from the critical point in which the Morse function increases \citep{2010arXiv1001.2933W}. For $n=1$, this means that the critical points of $h({\bm z})$ are saddle points. In general, what this means is that every critical point ${\bm z}_i$ for $h({\bm z})$ has associated with it a surface of steepest descent, $\mathscr{J}_i$, with dimension $n$, and a surface of steepest ascent, $\mathscr{U}_i$, of equivalent dimension. The surfaces of steepest descent are determined by the flow equations
\begin{equation}\label{eq:gradflow}
    \frac{dz^i}{dt} = \pm g^{ij}\frac{d h}{dz^j},
\end{equation}
for coordinates $z^i$ and spatial metric $g^{ij}$ for the complex plane. When the sign in front of the right-hand side of Eq.~\ref{eq:gradflow} is positive, the equation determines the trajectories of steepest ascent. Likewise, when the sign is negative, the equation determines the trajectories of steepest descent. The surface $\mathscr{J}_i$ associated with $z_i$ is a manifold determined by the set of initial points for which solutions to the steepest ascent equation terminate at $z_i$ as $t \to \infty$. Hence, the surface $\mathscr{J}_i$ is a surface of steepest \textit{descent}, since every direction pointing away from the critical point $z_i$ along the surface $\mathscr{J}_i$ is a direction of decreasing $h$. The dimension of $\mathscr{J}_i$ is the number of linearly independent directions away from $z_i$ in which $h$ decreases, which is the Morse index $n$. In general, one is free to choose the metric on the complex plane. For simplicity, we associate $\mathbb{C}^n$ with $\mathbb{R}^{2n}$ and apply the usual Euclidean metric.

One can show that along each of the surfaces of steepest descent, $\mathscr{J}_i$, the quantity $\mathrm{Im}\{ iS({\rm z}) \}$ is constant, and the total integrand is exponentially suppressed away from the critical point. This means that the integrand is localized and non-oscillatory along these surfaces, which we call the ``Lefschetz thimbles". Picard-Lefschetz theory tells us that the integral along the original domain, $\mathbb{R}^n$, is equal to a sum of integrals along the Lefschetz thimbles. Picard-Lefschetz theory also describes a means for determining which thimbles contribute to the integral: a thimble $\mathscr{J}_i$ contributes to the integral if and only if the contour of steepest ascent $\mathscr{U}_i$ for the corresponding critical point $z_i$ intersects the original integration domain. Numerically, finding the correct contour of integration is a simple matter. Any given point on the real plane will approach the relevant Lefschetz thimbles if allowed to evolve according to steepest descent equation (Eq.~\ref{eq:gradflow} with a negative sign). Thus, in order to find a contour of integration in which the integrand is localized and non-oscillatory, one must simply flow the real plane along the direction of steepest descent of the Morse function. See \citet{2010arXiv1001.2933W} for a more detailed description of the theoretical foundations of Picard-Lefschetz theory. \citet{job_pl} also gives a description of Picard-Lefschetz theory as applied to diffraction integrals. In Appendix~\ref{sec:plind}, we describe the specific numerical implementation of these ideas used in this paper.

Using Picard-Lefschetz theory, we can separate the total integral into contributions associated with a discrete set of critical points, $\{ {\bm z}_i \}$, by integrating along the associated Lefschetz thimbles separately. When $S({\bm z})$ is analytic, it obeys the Cauchy-Riemann equations, and therefore the critical points of $h$ are also the critical points of $S$. In the case of diffraction integrals in lensing, this means that the geometric images are also critical points of the Morse function. Moreover, the associated Lefschetz thimbles always contribute to the total integral, since the geometric images are just critical points for which ${\bm z_i} \in \mathbb{R}^n$ and, thereby, the contours of steepest ascent always intersect the real plane. The diffraction integral, as described by Picard-Lefschetz theory, is thus a clear extension of the Eikonal approximation. Instead of being a sum of the value of the integrand at a set of isolated points, the result is a sum of the integral performed over a discrete set of thimbles, where the integral is localized to regions near the critical points. This integrand becomes increasingly localized in the high-frequency limit. Note, however, that in general $h({\bm z})$ will have complex critical points as well, which may also contribute to the total integral.

\begin{figure}
    \centering
    \includegraphics[width=\columnwidth]{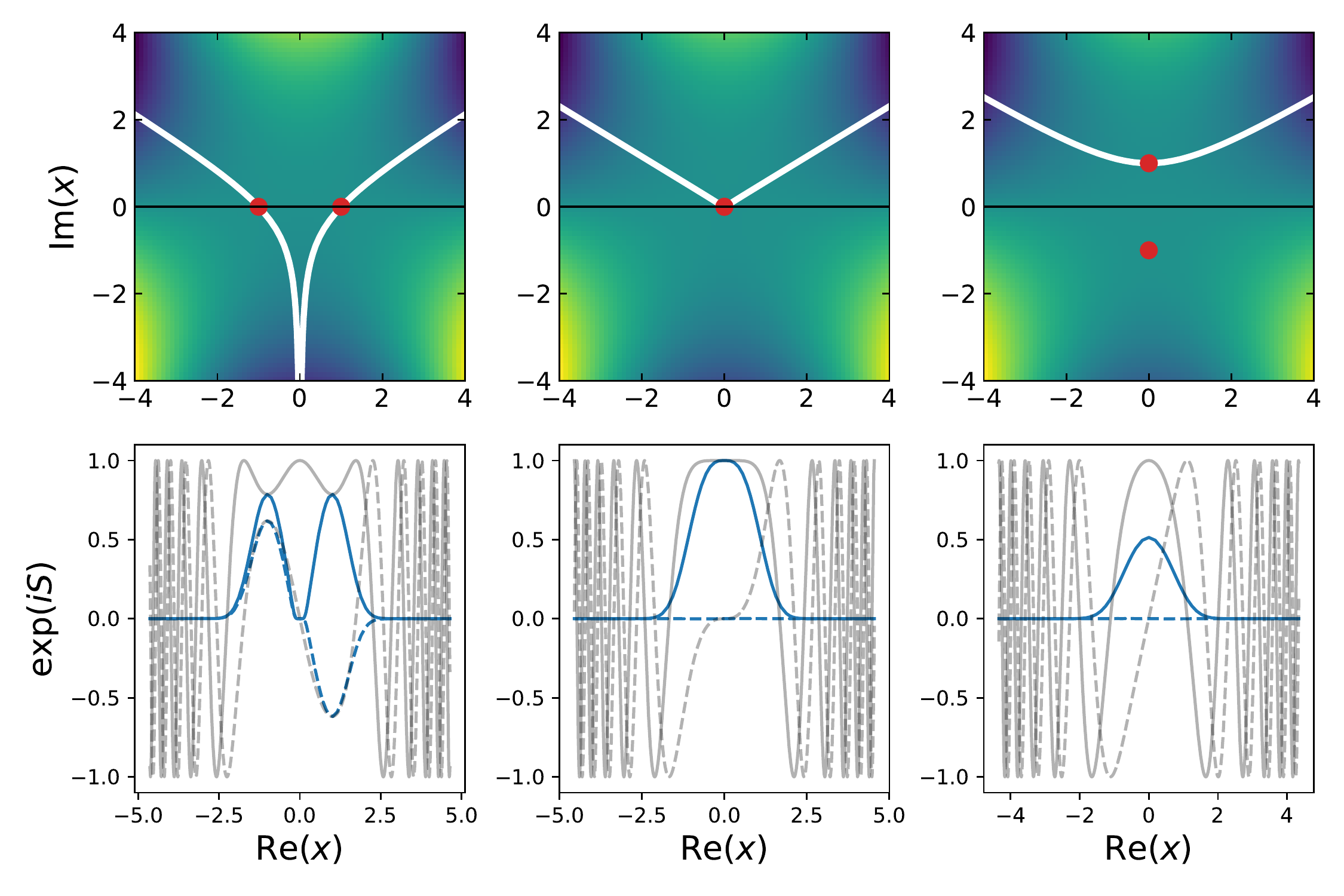}
    \caption{The top row shows the Lefschetz thimbles (in white) for the Airy function for different values of the parameter $y$. The left plot corresponds to $y = -1$, the middle plot to $y=0$, and the right plot to $y=1$. The red dots are the critical points, and the background colour shows the value of the Morse function, $h(x) = \mathrm{Re}\,iS(x)$ in the complex plane. The bottom row shows the value of integrand $e^{iS}$, evaluated along the Lefschetz thimbles (in blue) compared to the integrand evaluated along the real axis (in grey) for the corresponding values of $y$. The solid and dashed lines correspond to the real and imaginary parts of the integrand.}
    \label{fig:Airy_thimbles}
\end{figure}

Here, we turn to an example to make our explanation more concrete. Consider the Airy integral, a commonly encountered oscillatory integral with a known solution. 
\begin{equation}
\text{Ai}(y) = \frac{1}{2\pi} \int_{-\infty}^{\infty}\exp\left[i\left(x^3/3 +yx\right) \right] dx
\label{eq:Airyint}
\end{equation}
    $$= \frac{1}{2\pi} \int \exp\left[(\frac{z_i^3}{3}-z_r^2z_i- y z_i)+i(\frac{z_r^3}{3}-z_r z_i^2+y z_r) \right] dz.$$
Here, the Morse function is given by $h(x,y)= z_i^3/3-z_r^2z_i- y z_i$, where $z_r=\text{Re}(z), z_i=\text{Im}(z)$. The second line of Eq.~\ref{eq:Airyint} is the analytic continuation of the integral into the complex plane, and can be evaluated over any complex contour, but will in practice be evaluated over the Lefschetz thimbles associated with the Morse function. The top row of Figure \ref{fig:Airy_thimbles} shows the critical points in red and the relevant Lefschetz thimbles in white, for $y = -1, 0,$ and\,1, found using the numerical code described in Appendix \ref{sec:plind}. While for $y = -1$ and\,$0$, the relevant thimbles are associated with real critical points, for $y=1$ an imaginary critical point becomes relevant. In fact, for $y=1$ the only relevant critical point is imaginary. The bottom row of Figure \ref{fig:Airy_thimbles} shows the integrand computed along the thimbles compared with the integrand computed along the real axis. Along the real axis, the integrand is oscillatory across the entire domain, whereas it is non-oscillatory and localized along the thimbles. This concrete example is meant to illustrates the technique with which new classes of oscillatory integrals become numerically tractable. 

\subsection{Imaginary Images and Stokes Phenomena}
\label{sec:imaginary_images}

In Section~\ref{sec:geom}, we assumed that the relevant images that contribute to the Eikonal limit were the real solutions to the lens equation. However, as we have seen from Picard-Lefschetz theory, when we extend the phase function to the complex domain, imaginary images can contribute, if the corresponding contour of steepest ascent intersects the real plane. Likewise, imaginary images must also be taken into account in the Eikonal, or stationary phase approximation, as well. The general result for the field due to an imaginary image, ${\bm z}_i$, in the Eikonal limit is \citep{GrilloCordes2018}
\begin{equation}
    F_i = \frac{1}{|\Delta_i|^{1/2}} \exp \Big\{ i \big[S({\bm z}_i, {\bm y}; \nu) + \frac{\pi}{2} - \frac{\arg(\Delta_i)}{2} \big] \Big\}.
\end{equation}
The condition for an imaginary image to contribute to the Eikonal approximation is the same as the condition for their associated Lefschetz thimbles to contribute to the Picard-Lefschetz integral \citep{Wright_1980}. Unfortunately, in general there is no local way to determine whether an imaginary image contributes to the diffraction integral, i.e., there is no way to determine if an imaginary image, ${\bm z}_i$, contributes by evaluating quantities at ${\bm z}_i$ only. Ultimately, whether the associated steepest-ascent contour, $\mathscr{U}_i$, intersects the real plane must be determined. 

Images such that $h({\bm z}_i) > 0$ can automatically be excluded, since the associated steepest-ascent contour cannot intersect the real plane where $h=0$ everywhere. Thus, $h({\bm z}_i) < 0$ is a necessary condition for an image to contribute, but it is not generally sufficient, and the steepest-ascent contours must be computed. Since the locations of the imaginary images vary continuously with the parameters of the lens equation, it is not always necessary to perform the full thimble analysis for every imaginary image satisfying $h({\bm z_i})<0$ to determine whether it contributes. For example, when crossing a fold caustic, two real images will merge at a point and then pick up small imaginary parts with opposite signs. One of these images always continues to contribute on the other side of the caustic, and this will be the image whose value $h({\bm z_i})$ is smaller.  Note, in general there may be other imaginary images far from the merger point that contribute, for which the steepest-ascent contours must be computed. 

\begin{figure}
    \centering
    \includegraphics[width=\columnwidth]{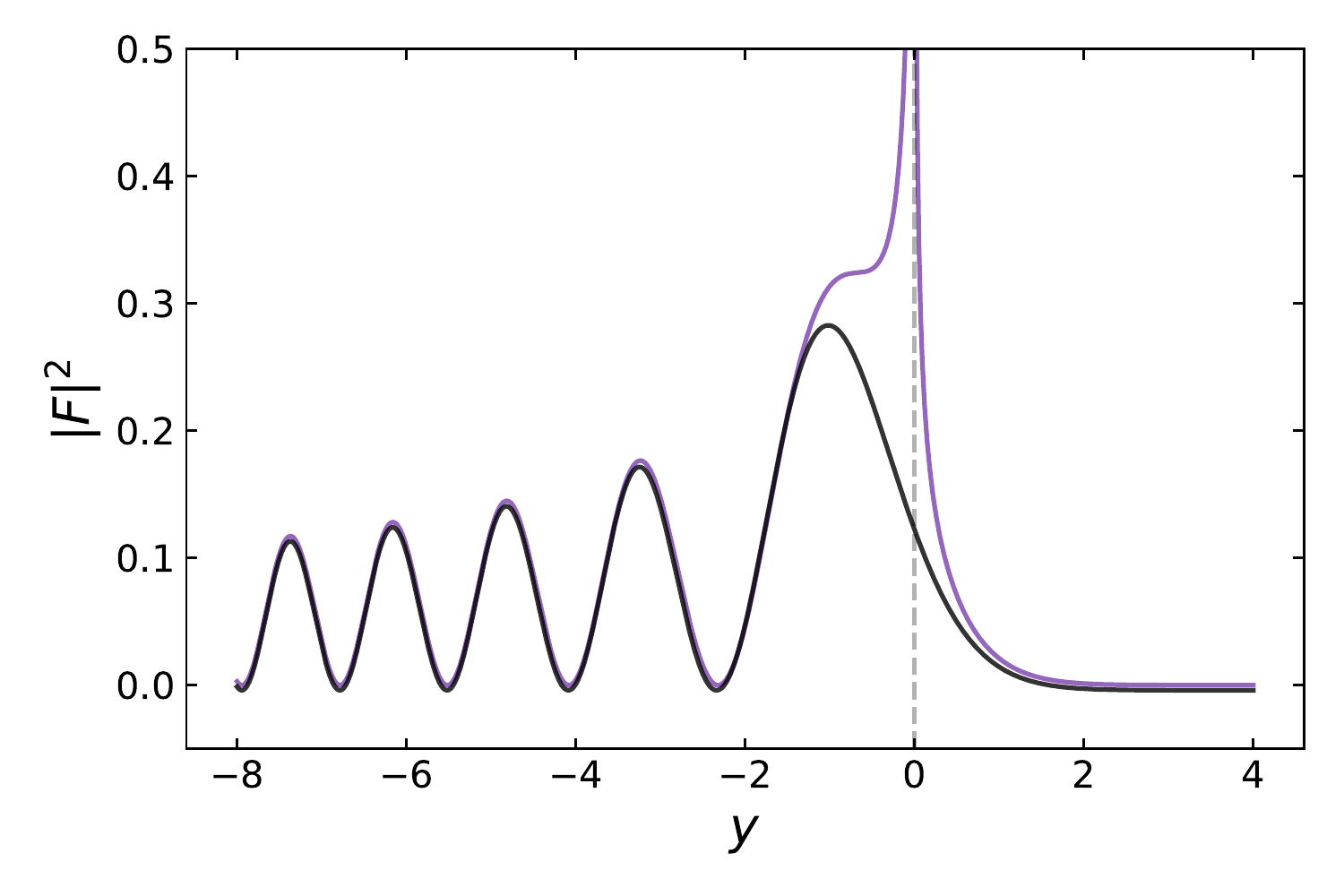}
    \caption{The square of the Airy function, $|\mathrm{Ai}(y)|^2$ (black), shown together with the full Eikonal approximation, taking into account imaginary images (purple). The dashed line at $y=0$ shows where caustic occurs, at which point the Eikonal approximation diverges.}
    \label{fig:Airy_eikonal}
\end{figure}

As a concrete example, consider the canonical form of the fold caustic which has the phase function
\begin{equation}
    S(x,y) = \frac{x^3}{3} + yx.
\end{equation}
This is the phase function of the Airy integral introduced in the previous section. The lens equation is $x^2 + y = 0$, which has solutions $x_\pm = \pm \sqrt{-y}$. For $y < 0$, there are two real solutions, and for $y > 0$ there are two imaginary solutions. The caustic occurs at $y=0$. The black curve in Fig.~\ref{fig:Airy_eikonal} shows the result of the integral, the Airy function. If one were to approximate the Airy function with the Eikonal limit, but only consider real images, then the approximation would be non-zero only when $y<0$, and would fail for positive $y$. Thus, at least one of the imaginary images must contribute. 

To see which images contribute, we show the relevant Lefschetz thimbles as $y$ goes from negative to positive in Figure \ref{fig:Airy_thimbles}. At the caustic, $y=0$, the images merge, and so have equal phase. Beyond the caustic, the imaginary images are complex conjugates, $x_{\pm} = \pm i \sqrt{|y|}$. Just beyond the caustic, the downward flow lines of the Morse function, $h$, will naturally only intersect the image $x_+$ as $\mathrm{Re} \{iS(x_+, y)\} < 0 < \mathrm{Re} \{iS(x_-, y)\}$. Thus, only $x_+$ is relevant. The purple curve in Fig.~\ref{fig:Airy_eikonal} shows the Eikonal approximation taking this into account. While we chose the Airy function for illustrative purposes, this behaviour is general for fold catastrophes, as the Airy phase function is simply the canonical form of the local expansion for fold catastrophes. Thus, when crossing a fold catastrophe, two real images merge and become imaginary; only the imaginary image with a smaller value of the Morse function will continue to contribute. 

What the Airy example shows is that, near caustics, it is clear the imaginary images must contribute, as at the caustic the images have formally infinite flux and they maintain that large flux even as they wander into the complex plane. But also, as we have seen, near the fold caustic it is relatively straightforward to diagnose which images continue to contribute. However, imaginary images can be important even far from caustics. In complex analysis, the term ``Stokes lines" refer to boundaries that separate a domain into regions based on the changing asymptotic behaviour of a function. In the context of lensing, Stokes lines are boundaries at which the number of relevant imaginary images changes by one. As \citet{Wright_1980} notes, this necessarily results in a jump discontinuity in the Eikonal, or stationary phase approximation of the diffraction integral, as at Stokes lines imaginary images can suddenly become relevant. This is similar to the case of caustics, in which the Eikonal approximation suffers an essential discontinuity (i.e. it diverges). At caustics, the number of relevant images also changes, but also the intensity of some number of those images diverge. Together, Stokes lines and caustics form boundaries that divide the source plane into regions with different numbers of relevant images. While Stokes lines do not correspond to regions where the intensity of images diverge, they indicate the presence of relevant imaginary images, which may have significant contribution to the total magnification for certain parameters.

Fig.~\ref{fig:pl_diagram} shows the Lefschetz thimble structure for the rational lens, which we will discuss in greater detail in Section~\ref{sec:rational}. The blue lines correspond to the Stokes lines for this lens. On one side of the Stokes lines, only a single real image contributes. This image continues to contribute as the Stokes line is crossed, but an additional imaginary image suddenly contributes. Though the number of contributing images changes across Stokes line (resulting in a discontinuity of the Eikonal limit), the diffraction integral is always analytic everywhere. The thimble diagrams in the circles of Fig.~\ref{fig:pl_diagram} represent what occurs on the corresponding Stokes lines or caustics. At the Stokes lines, the thimble for the real image terminates at the imaginary image, as opposed to elsewhere in parameter space, where it terminates at infinity or a pole. At the caustics, two of the real images merge to form a single critical point with higher multiplicity.

The importance of the contribution of imaginary images (sometimes referred to as ``complex rays") near caustics is well known, and has been stressed by multiple sources \citep[see e.g.][]{Keller:62, Wright_1980, nakamura, GrilloCordes2018}. However, the contribution of imaginary images near Stokes lines has not been widely considered in the lensing literature, although the importance of Stokes phenomena in optics has been known since the 80s \citep{Wright_1980}. Indeed, recent literature utilizing asymptotic methods for computing diffraction integrals have ignored Stokes lines, simply adding the contributions of all imaginary images satisfying $h<0$ \citep[see e.g.][]{nakamura, GrilloCordes2018}. While including irrelevant imaginary images results in only small errors in strong lensing regimes where there are multiple, bright real images, as \citet{Wright_1980} notes, in weak lensing regimes this can result in qualitatively different behaviour such as wave dislocations which are not present in the actual diffraction integral. 

In the next section, we will study the rational lens in detail to argue that these Stokes lines are of interest to the study of lensing. In particular, while caustics occur when many images become bright and merge, Stokes phenomena can occur in single (real) image regions. Whereas interference effects are readily apparent near caustics, the existence of interference effects in these weak-lensing regimes has not been widely discussed. But, as \citet{Jow2020} argue, frequency dependent interference effects can help break degeneracies when inferring physical parameters from lensing observations. Understanding interference effects in weak-lensing regimes may provide useful information even when dramatic changes in intensity due to caustics are not observed. 

\section{The Rational Lens}
\label{sec:rational}

We will now study a particular lens model to illustrate some of the concepts in the previous section. The lens model we will study is the 1D rational lens, given by
\begin{equation}
    S(x,y; \alpha) = \nu \Big[ (x-y)^2 + \frac{\alpha}{1+x^2} \Big],
    \label{eq:rat_S}
\end{equation}
where $\alpha > 0$ is the lens strength. For a plasma lens, $\alpha$ will be proportional to $\nu^{-2}$, while for a gravitational lens $\alpha$ will be completely independent of $\nu$. For the sake of generality, in this section we will treat $\alpha$ as independent. 

The lens mapping is given by
\begin{equation}
    y = \xi(x) = x - \frac{\alpha x}{(1+x^2)^2}.
    \label{eq:rat_lensmap}
\end{equation}
For a fixed $y$, the images are the solutions to a fifth-order polynomial equation:
\begin{equation}
    x^5 - y x^4 + 2 x^3 - 2yx^2 + (1-\alpha) x - y = 0.
    \label{eq:rat_lenseqn}
\end{equation}
This polynomial has five solutions, with at most three real solutions. We can divide the parameter space spanned by the parameters $y$ and $\alpha$ into distinct regions with Stokes lines as boundaries, as we describe in Section~\ref{sec:imaginary_images}. In each region, a different number of images contribute to the diffraction integral. Analyzing the Lefschetz thimbles associated with each image, we determine which images contribute and which are irrelevant. 

\begin{figure}
    \centering
    \includegraphics[width=\columnwidth]{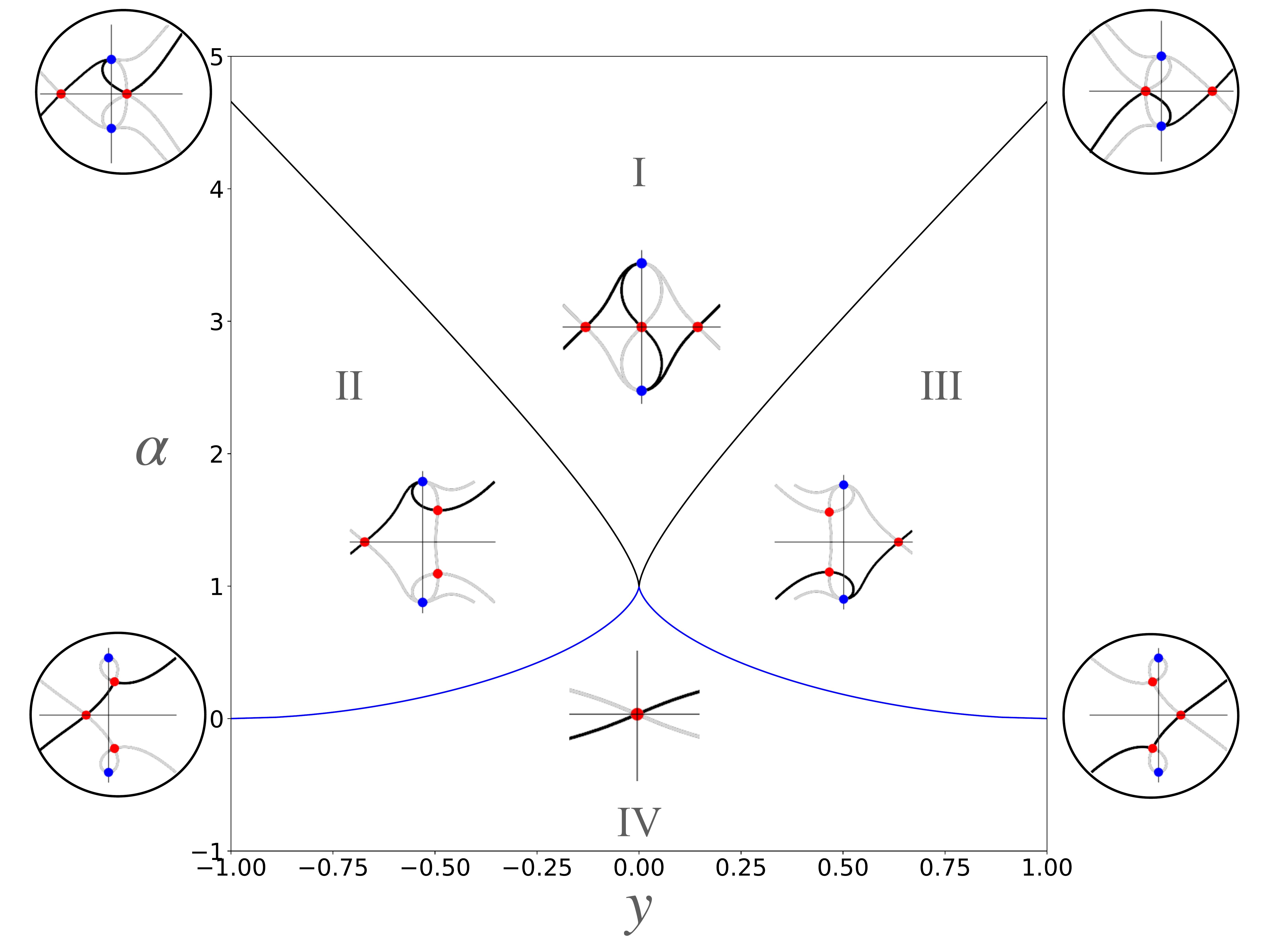}
    \caption{The parameter space of the 1D rational lens spanned by the parameters $\alpha$ and $y$, divided into regions bounded by caustics and Stokes lines. The black curve corresponds to caustics, and the blue curve corresponds to the Stokes lines. In each region, a representative plot of the Lefschetz thimble structure for the lens is shown, i.e. a plot in the complex plane of the steepest descent and ascent contours of the Morse function, $h(x) = \mathrm{Re}\,iS(x)$, emanating from the critical points (shown in red) of the Morse function. The Lefschetz thimbles are the steepest descent contours, shown in black, and the steepest ascent contours are light-grey. The blue dots correspond to the poles of the Morse function, and are located at $x = \pm i$. The topology of the Lefschetz thimbles only changes between the regions labelled with Roman numerals, and the thimble structure on the boundaries are shown in the circled diagrams at the corners of the figure. The parameters $\alpha$ and $y$ are dimensionless parameters which correspond to lens strength and source position. In plasma lensing $\alpha$ is proportional to inverse frequency squared.}
    \label{fig:pl_diagram}
\end{figure}

Fig.~\ref{fig:pl_diagram} shows the different regions bounded by Stokes lines for the rational lens, as well as diagrams of the thimble structure associated with each region. Although Eq.~\ref{eq:rat_lenseqn} has five solutions, at most three solutions contribute to the diffraction integral. This occurs in Region I where the three relevant images are real solutions to Eq.~\ref{eq:rat_lenseqn}. As one crosses from Region I to Regions II and III, two of the real images merge and become imaginary, with only one of the resulting imaginary images remaining relevant in Regions II and III. Thus, the boundaries separating Regions I, II, and III are, in fact, caustics. Region IV is a region in which only one image is relevant, and this is a real image. As one crosses from Region IV to Regions II and III, the real image remains relevant, but the thimble topology changes across the Stokes lines so that an imaginary image becomes relevant. Note that Fig.~\ref{fig:pl_diagram} only shows the $(y, \alpha)$-parameter space, and does not include the parameter $\nu$. This is because $\nu$ is simply an overall pre-factor in Eq.~\ref{eq:rat_S}, and so does not change the steepest-descent or ascent contours of the Morse function (i.e., it does not change the thimble structure). 

To find the location of the caustics one can compute the Jacobian of the lens mapping
\begin{equation}
    \Delta = \frac{\partial \xi}{\partial x} = 1 + \frac{ \alpha (3x^2 - 1)}{(1+x^2)^3}.
\end{equation}
The set of $x_i$ where $\Delta = 0$ corresponds to the critical curves, which, when mapped to the source plane, $y_i = \xi(x_i)$, give the caustic curves in the source plane. 

The location of the Stokes lines (the blue lines in Fig.~\ref{fig:pl_diagram}) are found by computing for what values of $y$ and $\alpha$ do any pair of solutions, $x_i$ and $x_j$, of Eq.~\ref{eq:rat_lenseqn} satisfy 
\begin{equation}
    {\rm Im}\, \{ iS(x_i, y; \alpha)\} = {\rm Im} \,\{ iS(x_j, y; \alpha)\}.
\end{equation}
This is a necessary condition for a Stokes line, since ${\rm Im} iS$ is constant along a given Lefschetz thimble. For the thimble of the real image to terminate at the imaginary image, the images must have equal values of ${\rm Im} iS$.

As shown in Fig.~\ref{fig:pl_diagram}, for $\alpha > 1$, at the boundaries between regions, two of the images merge to form a single image of higher multiplicity on the real line. Thus, these boundaries correspond to singularities in the lens mapping. That is, for $\alpha > 1$ the boundaries between regions correspond to caustics, where the geometric and Eikonal magnifications are formally infinite. In contrast, for $\alpha < 1$, the boundaries between regions are Stokes line. At Stokes lines, the Lefschetz thimbles for two images coincide, but the images themselves remain distinct. The point $\alpha = 1$ and $y=0$ is a special point, and is known as a cusp catastrophe, where all three relevant images merge at a point. Here also, the caustics and Stokes lines intersect.

\subsection{Stokes lines in the rational lens}
\label{sec:stokes}

The central question of this paper is whether or not wave effects can be observed in regions where there is only a single real image. In other words, can interference between real and imaginary images be observed in the single real-image regime, often referred to as the ``weak lensing" regime ($\alpha$<1). We will explore this question by studying the contribution of imaginary images in wave optics near Stokes lines in the rational lens. 


\begin{figure}
    \centering
    \animategraphics[label=stokes, controls, timeline=Figures/rational_lens/timeline.txt,width=\columnwidth]{6}{Figures/rational_lens/frame_}{000}{042}
    \caption{(Top panel) Embedded animation of the Lefschetz thimbles (open in Adobe Reader to view animation), coloured to their phase contribution to the diffraction integral, for the rational lens ($\alpha = 0.9, \nu=10$) for different values of $y$, with the default frame set at $y=-0.02$. Dots show the real image and the relevant imaginary image, coloured to their phases computed in the Eikonal regime. The size of the dots is proportional to their intensities, and the imaginary image is shown as a ``$\times$'' when it's too small to be displayed. The background colour shows the value of the Morse function. The shaded region is below the threshold value (See Appendix~\ref{sec:plind}). The white dashed line indicates the real line.
    (Bottom panel, top) The position, $x$, in the lens plane of the relevant images for different values of $y$, where the imaginary image is relevant only when $|y| > 0.014$. The absolute value of the position of the imaginary image multiplied by the sign of its real part is shown in purple. The dots are the same as in the top panel.
    (Bottom panel, middle) Amplitude, $F_i$, of each individual image computed in their respective regime. The solid lines are the real part and the dashed lines are the imaginary part of $F_i$.
    (Bottom panel, bottom) Total intensity of each regime.}
    \label{fig:stokes_thbt}
\end{figure}

Fig.~\ref{fig:stokes_thbt} shows a comparison for the geometric, Eikonal, and full wave regimes for the rational lens, for $\alpha = 0.9$ and $\nu = 10$. For this choice of $\alpha$, the Stokes lines occurs at $y = \pm y^* \approx \pm 0.014$. When $|y| < y^*$, only one image is relevant, and this a real image. When $|y| > y^*$ an imaginary image becomes relevant. The top panel of Fig.~\ref{fig:stokes_thbt} shows an animation of how the thimble structure changes as $y$ changes, showing how the imaginary images go from relevant to irrelevant across the Stokes lines. Note that for the geometric optics regime, only the real image is ever relevant. For the Eikonal and full wave optics regimes, an imaginary image becomes relevant when $|y| > y^*$. Analytically, except for when $|y| = y^*$ exactly, one end of the Lefschetz thimbles associated with both the real and imaginary images always terminates at a pole. Note, however, in the animation, for certain values of $y$ near the Stokes lines the thimbles appear to terminate at a point away from the pole. This is simply a numerical artefact due to having imposed a maximum integration time for solving the flow equation. This does not affect the numerical result for the diffraction integral, as long as the integration contour is close to the true Lefschetz thimbles.

The top of the bottom panel of Fig.~\ref{fig:stokes_thbt} shows the position, $x$, in the lens plane of the real image as a function of $y$ in black. The purple curve shows the absolute value of the position of the imaginary image, multiplied by the sign of its real part, for the values of $y$ for which it is relevant. The middle of the bottom panel is the field, $F_i$, computed for the different images in the geometric, Eikonal, and wave regimes. The solid lines correspond to the real part of $F_i$, and the dashed lines correspond to the imaginary part. Since only the real image is relevant in the geometric case, and since geometric optics does not take into account wave effects, there is only one curve for the geometric case, which is the square-root of intensity of the real image, shown in black. This curve acts as an envelope for the value of $F_i$ of the real image in the Eikonal limit. In addition to the real image, the Eikonal limit also has a contribution from an imaginary image when $|y| > y^*$, which is shown in purple. Shown in red and brown, are the corresponding curves for the real and imaginary images computed in the full wave regime using Picard-Lefschetz theory. This comparison between the fields computed for each image \textit{individually} across the three regimes is made possible by Picard-Lefschetz theory, as it gives us the ability to assign contributions to the total diffraction integral to individual images, by associating the contribution from individual thimbles to their corresponding image. In contrast, methods for evaluating the diffraction integral that perform the integration over the entire real domain do not have the ability to separate contributions in this way.

The bottom of the bottom panel of Fig.~\ref{fig:stokes_thbt} shows the total intensity, $|\sum F_i|^2$, for the different regimes. Since, for the geometric limit, there is only one image, this is the same as the intensity for the real image. For the Eikonal limit, there is a sharp discontinuity at $y = \pm y^*$. Since the intensity of the individual images in the Eikonal limit is continuous, but the imaginary image suddenly becomes relevant at the Stokes line, this leads to a discontinuity in the total intensity. This discontinuity is a result of the breakdown of the Eikonal approximation at Stokes lines, rather than a physical effect. As discussed in Sec.~\ref{sec:imaginary_images}, these discontinuities at Stokes lines are a generic feature of the Eikonal, or stationary-phase approximation. Since the individual images always have continuous intensity in the Eikonal limit (except at caustics where they diverge), the total intensity will always be discontinuous at Stokes lines, where images suddenly become relevant. In contrast, the full wave intensity does not exhibit a discontinuity at the Stokes line, as the diffraction integral is analytic, meaning the total intensity must be continuous everywhere. However, in order to maintain continuity of the total intensity, the intensities of the individual images must be discontinuous at the Stokes lines in the full wave regime. Another important difference to note between the Eikonal and full wave results, is the fact that for the full wave regime, near the Stokes line, the imaginary image actually has a greater intensity than the real image. Away from the Stokes line, the intensity of the imaginary image quickly drops, and the Eikonal limit becomes a better approximation. Since we are in a weak lensing regime, geometric optics also becomes a better approximation as the intensity of the imaginary image gets smaller.

The qualitative picture sketched here is the same for all values of $0 < \alpha < 1$, but the locations of the Stokes lines are different, and as $\alpha$ decreases, and we move further away from the cusp in parameter space, the intensities of all the images involved decrease. When one is very far from the cusp, the contribution from the imaginary image is essentially negligible, and the total intensity curve for the three regimes all coincide. 

\subsection{Imaginary images in the single real image regime}
\label{sec:imaginary_single}

Geometric optics is an inadequate description of the lensing of coherent sources as it fails to take into account wave effects. In the weak-lensing regime, ($\alpha < 1$), the deviations of wave optics from geometric optics is partially a result of the interference between the real image and imaginary image that become relevant near Stokes lines (the other cause being proximity to the cusp, as at the cusp the geometric magnification diverges, whereas magnifications are always finite in wave optics). The size of the contribution of the imaginary image to the total intensity depends on the lensing parameters, and may be exponentially small. In these cases, the interference pattern will be effectively unobservable. Here we use the term ``interference pattern" to refer to the oscillatory behaviour of the intensity in the wave regime. These oscillations are the result of the presence of imaginary images, in addition to the single, real image, and produce a qualitatively different intensity curve than geometric optics. For the rational lens, the geometric optics intensity monotonically decreases with increasing $|y|$ in the weak lensing regime, and therefore exhibits no oscillatory behaviour.

\begin{figure}
    \centering
    \includegraphics[width=\columnwidth]{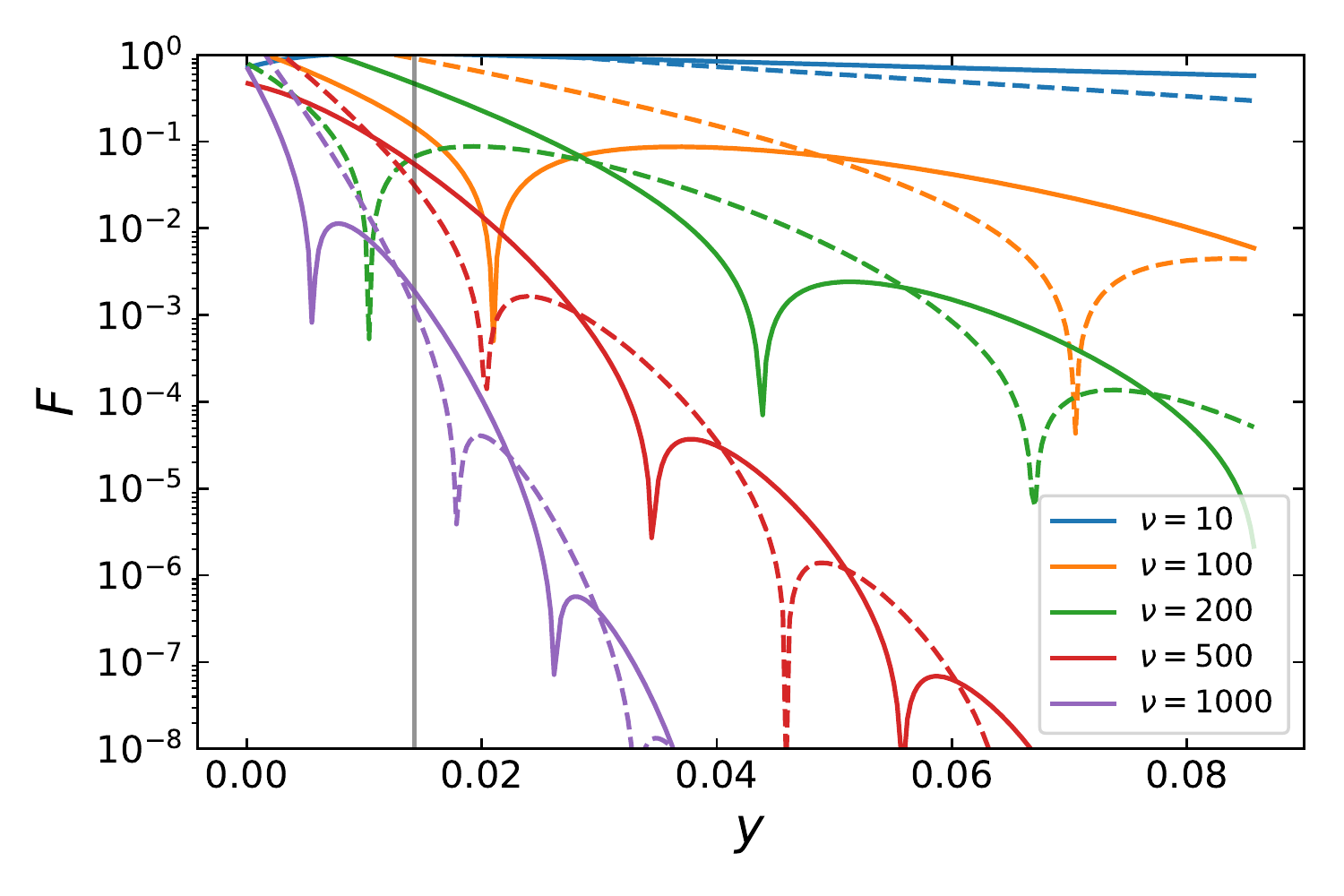}
    \caption{The complex wave field, $F_i$, of the imaginary image computed in the Eikonal limit for the rational lens, with $\alpha = 0.9$. The field is computed as $y$ crosses the Stokes line (the grey vertical line at $y \approx 0.014$) for various values of $\nu$. The solid and dashed lines correspond to the real and imaginary parts, respectively, of the complex field, $F_i$.}
    \label{fig:imag_eik}
\end{figure}

To diagnose when the imaginary image becomes significant we can look at how the field, $F_i$, of the complex image varies with the lens parameters in the Eikonal limit. Although, as we have discussed, the Eikonal limit fails at the Stokes lines due to its discontinuity, it provides a convenient way to determine the relative scale of the imaginary image's contribution. Fig.~\ref{fig:imag_eik} shows how the Eikonal limit of the imaginary image scales with the frequency parameter $\nu$ as one crosses a Stokes line for $\alpha = 0.9$. The intensity decays exponentially as a function of $y$ as one moves away from the Stokes line, so that the imaginary image is brightest at the Stokes line. As a function of frequency, this amplitude is largest for lower frequencies. This is consistent with the generic statement that the Eikonal approximation is valid in the limit $\nu \to \infty$, as the Eikonal limit, near Stokes lines, is a good approximation of the full wave optics when the imaginary image is negligible. For low frequency, e.g. $\nu = 1$, the intensity of the imaginary image can be of order unity relative to the real image, which may be significant.

\begin{figure}
    \centering
    \includegraphics[width=\columnwidth]{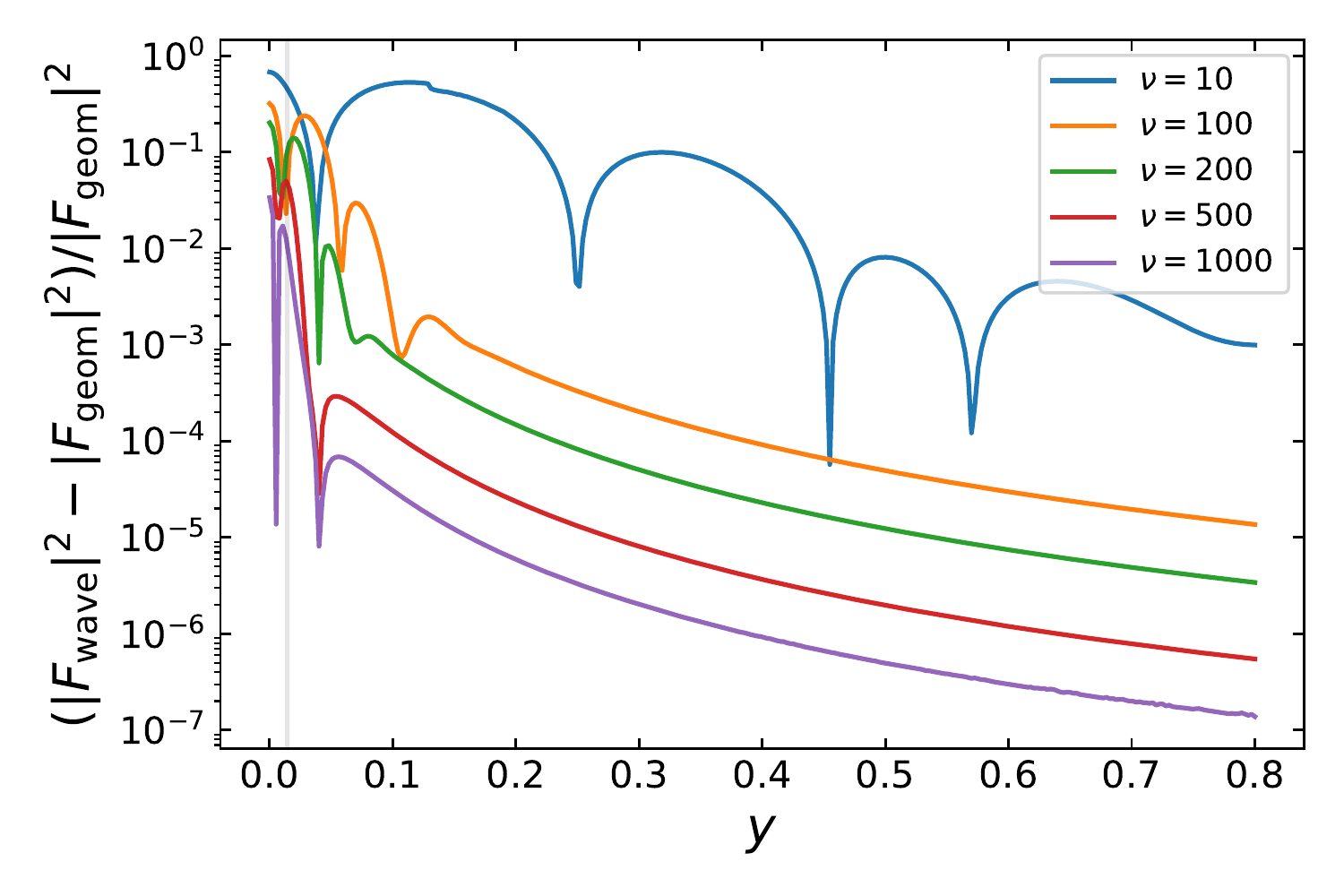}
    \caption{The fractional deviation of the full wave optics intensity from the geometric optics intensity for the rational lens with $\alpha = 0.9$. The intensities are computed as $y$ crosses the Stokes line (the grey vertical line at $y \approx 0.014$) for various values of $\nu$. We refer to this deviation from geometric optics as wave interference effects. }
    \label{fig:wavevgeom}
\end{figure}

Now, interference patterns are characterized by the oscillatory nature of the total intensity. While the oscillations may be difficult to distinguish by eye in the bottom panel of Fig.~\ref{fig:stokes_thbt}, we can compute the fractional deviation of the wave optics intensity from the geometric optics result. This is shown in Fig.~\ref{fig:wavevgeom} for $\alpha = 0.9$, and for a variety of frequency parameters. This fractional difference reveals the oscillations caused by interference with the imaginary images, but for high frequencies the amplitude of the oscillations are small. However, for $\nu = 100$ and lower frequencies, the amplitude can be made significantly larger than percent level, implying that these oscillations could, in principle, be observable.  

In the Eikonal limit, the interference pattern begins at the Stokes line, as prior to the Stokes line the imaginary image does not contribute, and thus cannot interfere with the real image. Thus, in the Eikonal limit, the Stokes line represents a boundary where the intensity behaves qualitatively different on either side. As discussed in the previous section, the discontinuity at the boundary is a mathematical artefact due to the breakdown of the stationary phase approximation at the Stokes line, rather than a physical effect. Indeed, in the full wave optics regime, the oscillations begin well before the Stokes line. Thus, in wave optics, there is no hard boundary at which an additional image can be said to start to interfere with the real image. Nevertheless, as $\nu \to \infty$, the region in which there are observable oscillations becomes increasingly localized to near the Stokes line. This is analogous to the situation near caustics: in geometric optics, the caustic represents a hard boundary between a bright region and a dark region. In wave optics, that boundary is smeared out by wave effects, especially for low frequencies.  

\begin{figure}
    \centering
    \includegraphics[width=\columnwidth]{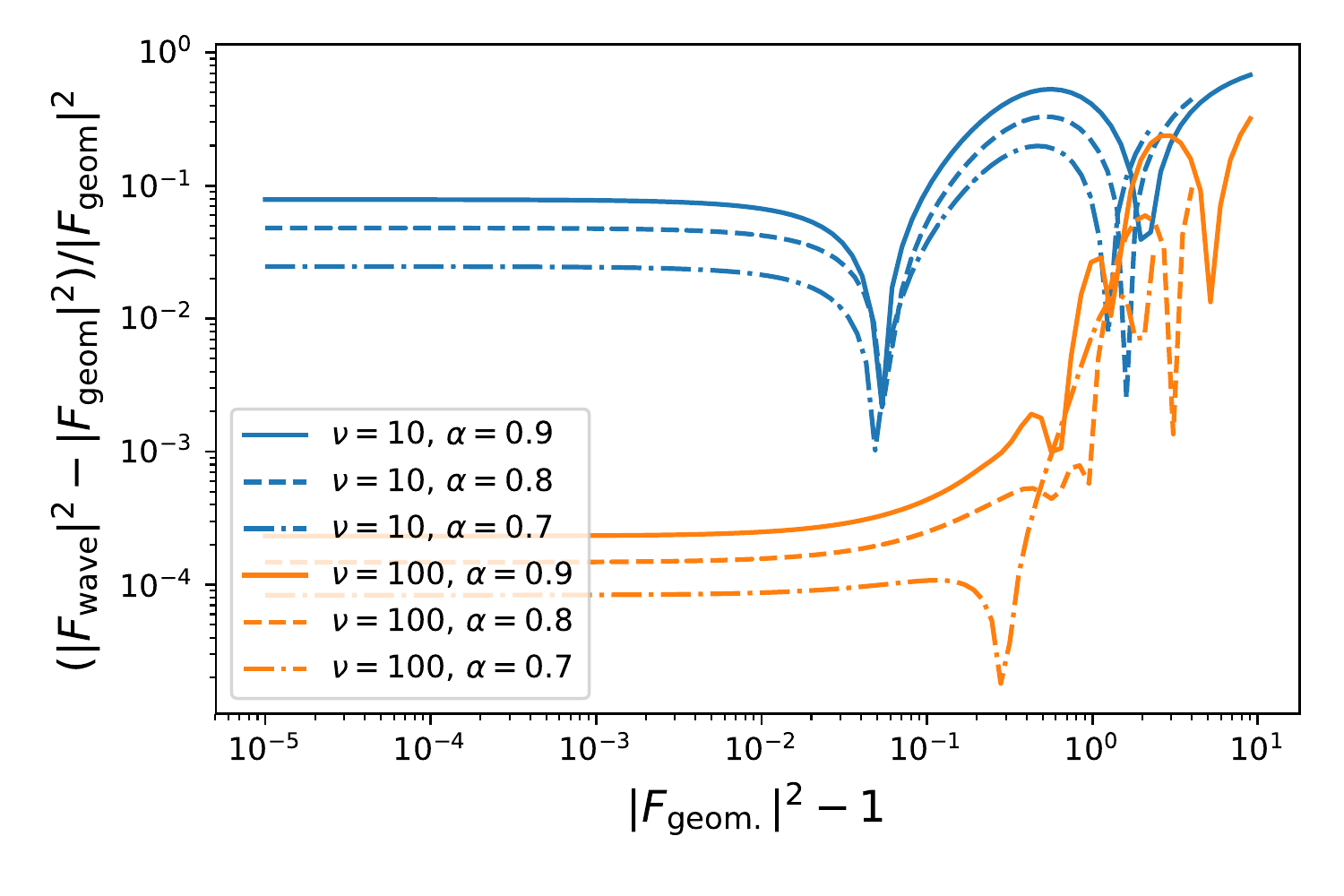}
    \caption{The fractional deviation of the full wave optics intensity for the rational lens from the geometric optics intensity computed as a function of the geometric intensity above unity. The curves are computed for various values of $\alpha$ and $\nu$.}
    \label{fig:stokes_geommag}
\end{figure}

Since the geometric optics intensity monotonically decreases with $|y|$, we can reparametrize the $x$-axis of Fig.~\ref{fig:wavevgeom} to depend on the geometric intensity: specifically, $|F_\mathrm{geom.}|^2-1$ instead of $y$. Doing so, we compute the deviation of the wave optics intensity from the geometric intensity as a function of the geometric intensity for various values of $\alpha$ and $\nu$, shown in Fig.~\ref{fig:stokes_geommag}. The lens strength, $\alpha$, determines the peak value of the geometric intensity, which corresponds to the leftmost value of $|F_\mathrm{geom.}|^2-1$ attained by the various curves in Fig.~\ref{fig:stokes_geommag}. Note that for low frequencies ($\nu = 10$), even when the geometric magnification is small, the deviation caused by the contribution of the imaginary image can be as large as ten percent.

We now have an answer to the central question posed at the beginning of Section~\ref{sec:stokes}. That is even in the weak-lensing regime, there exist regions of parameter space in which interference patterns are potentially significant, due to the contribution of imaginary images. As Figs.~\ref{fig:wavevgeom} and \ref{fig:stokes_geommag} show, the interference pattern depends both on the parameters $\alpha$ and $\nu$, independently. This is a crucial point to note, as this means that even in the weak-lensing regime phase information can be extracted from a lensing event.

The spacing between the oscillations in $y$ is effectively determined by the frequency $\nu$. When the source is located far from the centre of the lens, i.e. when $y$ is large, the relative phase between interfering images is dominated by the geometric term, which goes like $\sim\nu y^2$. Thus, for large $y$, the spacing between oscillations in intensity as a function of $y$ roughly scales as $\sim 1/\sqrt{\nu}$, independent of the precise form of the lensing potential. In this case, a measurement of an interference pattern can provide a rough, order-of-magnitude estimate of $\nu$, irrespective of the details of the lens model. In general, of course, a precise measurement of $\nu$ will require fitting the observed intensity to a particular lens model. In either case, this is in contrast to geometric optics, where the total magnification depends only on the lens mapping, Eq.~\ref{eq:rat_lensmap}, which does not depend on $\nu$ independently of $\alpha$. 

More detailed phase information can be gained by looking at the oscillations in $\nu$. Generally, lensing observations are taken at multiple frequencies, and so one observes the same lensing event at different values of $\nu$. To illustrate, we will use the Eikonal limit for simplicity, as it is straightforward to compute the phase of the images. The oscillations in $\nu$ are determined by the phase difference, $\delta_{xz} = \mathrm{Im}\{ iS(x, y; \nu) - iS(z, y; \nu)\}$, where $x$ is the real image, and $z$ is the imaginary image. The frequency of the oscillations in $\nu$ are given by $\frac{d \delta_{xz}}{d \nu}$, which is the difference in group delay between the two images. For a plasma lens, where $\alpha \propto \nu^{-2}$, the group delay of a single image is given by
\begin{equation}
    \frac{d (-\mathrm{Im}iS)}{d \nu} =  \mathrm{Re} \big\{(x-y)^2 - \frac{\alpha}{1+x^2} \big\}.
\end{equation}
The Fourier transform of the intensity for a narrow bandwidth, $\Delta \nu$, for a fixed $y$ yields a delta function centred at the differential group delay $\delta_{xz}$ in the Eikonal limit. The requirement that the bandwidth be narrow is so that the image positions $x$ and $z$, and the value of $\alpha$ are roughly constant with respect to $\nu$. In the full wave optics regime, the result is similar, but the delta function is smeared out, especially at low frequencies. Thus, a measurement of the interference pattern in $\nu$ allows one to directly measure the differential group delay of the images. In gravitational lensing, where $\alpha$ is independent of $\nu$, the group delay is equal to the phase delay. Measuring the delay between the images is not possible in geometric optics, unless the time of arrival between each image is large enough that the images arrive at the observer separated in time (as, for example, in the case of FRB microlensing \citep[see e.g.][]{munoz_lensing_2016}.

It has been widely assumed that in the weak-lensing regime, since there is only a single geometric image which cannot interfere with itself, it is not possible to retrieve this kind of phase information. Indeed, if one neglects the contribution from imaginary images, the Eikonal limit reduces to geometric optics, which contains no phase information. However, what we have shown in this section is that even in the weak-lensing regime, the contribution from imaginary images leads to potentially observable interference patterns.

\subsection{Physical parameters}
\label{sec:phys}


Ultimately, the goal of modeling lensing events is to extract the underlying physical parameters of the lensing system. So far, for the sake of generality, we have examined the rational lens using dimensionless variables, $\nu$, $\alpha$, $y$, and $x$. Using Eq.~\ref{eq:F}, we can write these out in terms of physical parameters. 

For a plasma lens, the lensing potential is given by $\psi(\theta, \omega) = \Sigma_e(\theta) e^2 / 2 m_e \epsilon_0 \omega^2$, where $\Sigma_e$ is the electron column density and $\omega$ is the (angular) frequency of observation. Thus, for a plasma lens in the form of Eq.~\ref{eq:rat_S}, the dimensionless parameters are given by
\begin{align}
    y &= \frac{\beta}{\theta_0}, \\
    \nu &= \frac{\omega}{2c} D \theta_0^2, \\
    \alpha &= \frac{2ck}{\omega^2} \frac{\Sigma_0}{D\theta_0^2} = \frac{kD\theta_0^2 \Sigma_0}{2c\nu^2},
\end{align}
where $\Sigma_0$ is the peak electron column density of the lens, $k = e^2 / 2m_e \epsilon_0 c$, and $\theta_0$ is a characteristic size of the lens chosen such that $x = \theta/\theta_0$.

From this, we can see the utility in modelling wave effects. A measurement of the intensity of a source undergoing a lensing event will only allow for the inference of these three dimensionless parameters, assuming the angular separation of the images are un-resolvable. As discussed previously, geometric optics is fully determined by the lens mapping (Eq.~\ref{eq:rat_lensmap}), and so only depends on $y$ and $\alpha$. Thus, in geometric optics, there is an entanglement between $\alpha$ and $\nu$, as the observed intensity will only depend on $\nu$ through the dependence of $\alpha$ on $\nu^{-2}$. This results in a three-way degeneracy between $\Sigma_0$, $D$, and $\theta_0$, as the geometric optics intensity depends only on the combination $\Sigma_0 / D \theta_0^2$. This degeneracy cannot be broken from a single observation of the source intensity during a lensing event. However, for coherent sources that exhibit wave effects, the observed intensity depends on $\nu$ independently of $\alpha$ (holding $\alpha$ fixed and varying $\nu$ yields a different intensity curve). Thus, $D \theta_0^2$ can be measured independently of $\Sigma_0$. As we have shown, these interference effects persist in the weak lensing regime, and so this degeneracy can be broken even with only a single real image.

Consider a measurement of a light curve from a point source being lensed, where the relative transverse angular velocity of the source and lens is $\mu_\mathrm{rel.}$. Then $y(t) = (t-t_0) \mu_\mathrm{rel.} / \theta_0$, where $t_0$ is the time when $y=0$. In this case, the underlying physical parameters that can be inferred from a measurement of $y(t)$, $\nu$, and $\alpha$ are $\Sigma_0$,  $D\theta_0^2$, and $\mu_\mathrm{rel.}^2/\theta_0$. Alternatively, one can rearrange these combinations to obtain $\Sigma_0$,  $D\theta_0^2$, and $\mu_\mathrm{rel.}^2 D$. Thus, the angular velocity, scale of the lens, and distance $D$ are not, in general, possible to disentangle from each other from a measurement of the intensity of a single lensing event alone. Also recall that the distance $D$ is itself an effective distance, and is a combination of the distances involved, $D = D_d D_s / D_{ds}$. However, in a variety of practical applications, one or more of these parameters are known from independent observations, thus allowing for a complete determination of the underlying physical parameters. For example, if the lens has a known visible counterpart, the distances and angular velocities of both source and lens may be known a priori.

As another example, consider the case where the lens is very close to the source; for example, in the case of the pulsar B1957+20, which undergoes extreme plasma lensing by the ionized outflow of its companion \citep{2018Natur.557..522M}. In this case, $D_{ds} \ll D_d \approx D_s$. Using this, we can rewrite the lensing parameters as
\begin{align}
    y &= \frac{\eta}{a} = (t-t_0)\frac{u}{a}, \\
    \nu &= \frac{\omega}{2c} \frac{a^2}{\overline{d}} \label{eq:nu}, \\
    \alpha &= \frac{2ck}{\omega^2} \frac{\Sigma_0 \overline{d}}{a^2} \label{eq:alpha},
\end{align}
where $a = \theta_0 D_d$ is the physical size of the lens, $\eta = \beta D_s$ is the physical position of the source, $u$ is the relative transverse velocity of the source and lens, and $\overline{d} = \frac{D_{ds} D_d}{D_s} \approx D_{ds}$. For PSR B1957+20, the distance between the pulsar and its companion is well-constrained, thus, a measurement of lensing event in the wave regime can be used to directly measure the relative transverse velocity, the size of the lens, and $\Sigma_0$.

\section{Observational prospects}
\label{sec:prospects}

So far, we have shown that, in wave optics, imaginary images can produce significant interference patterns, even in weak-lensing regimes. The presence of wave effects means that phase information can be extracted, which, naively, one might think would not be possible in a weak-lensing regime. In this section, we will briefly discuss the possibility of detecting these interference patterns in single-image lensing events in observations.

First, we will remark that while we have focused on the rational lens, the presence of Stokes lines and imaginary images is generic for a wide variety of lens models. The rational lens was chosen as an example to illustrate the effects of imaginary images for its simplicity, and because it is a rational approximation of the Gaussian lens, which has been widely studied and applied to data \citep[see e.g.][]{1998ApJ...496..253C, 2017ApJ...842...35C}. The rational and Gaussian lenses produce similar lens mappings, and therefore, produce a qualitatively similar caustic structure, and, in this sense, are good approximations of one another. However, one may ask whether or not the qualitative picture we have outlined for imaginary images is strongly dependent on the precise details of the lensing potential.

To investigate this question, we consider the Gaussian potential, which is widely used in the literature \citep[see e.g.][]{1998ApJ...496..253C, 2012MNRAS.421L.132P, 2015ApJ...808..113C}. Since the lens equation for the Gaussian lens is not a simple polynomial and has infinite imaginary roots, the diffraction integral is not trivial to compute, even with Picard-Lefschetz theory. Thus, instead, we compute the diffraction integral for a lens with potential $\hat{\psi} = \alpha/(1+x^2+nx^4)$, where $n \approx 0.53$. This inverse-quartic lens and the simple rational lens considered in Section~\ref{sec:rational} are both the simplest rational approximations of a given order to a Gaussian lens, minimizing the average residual with the Gaussian lens while keeping curvature at the origin fixed. If the rational lens is considered an approximation to the Gaussian lens, this inverse-quartic lens is the next-order approximation.

\begin{figure}
    \centering
    \includegraphics[width=\columnwidth]{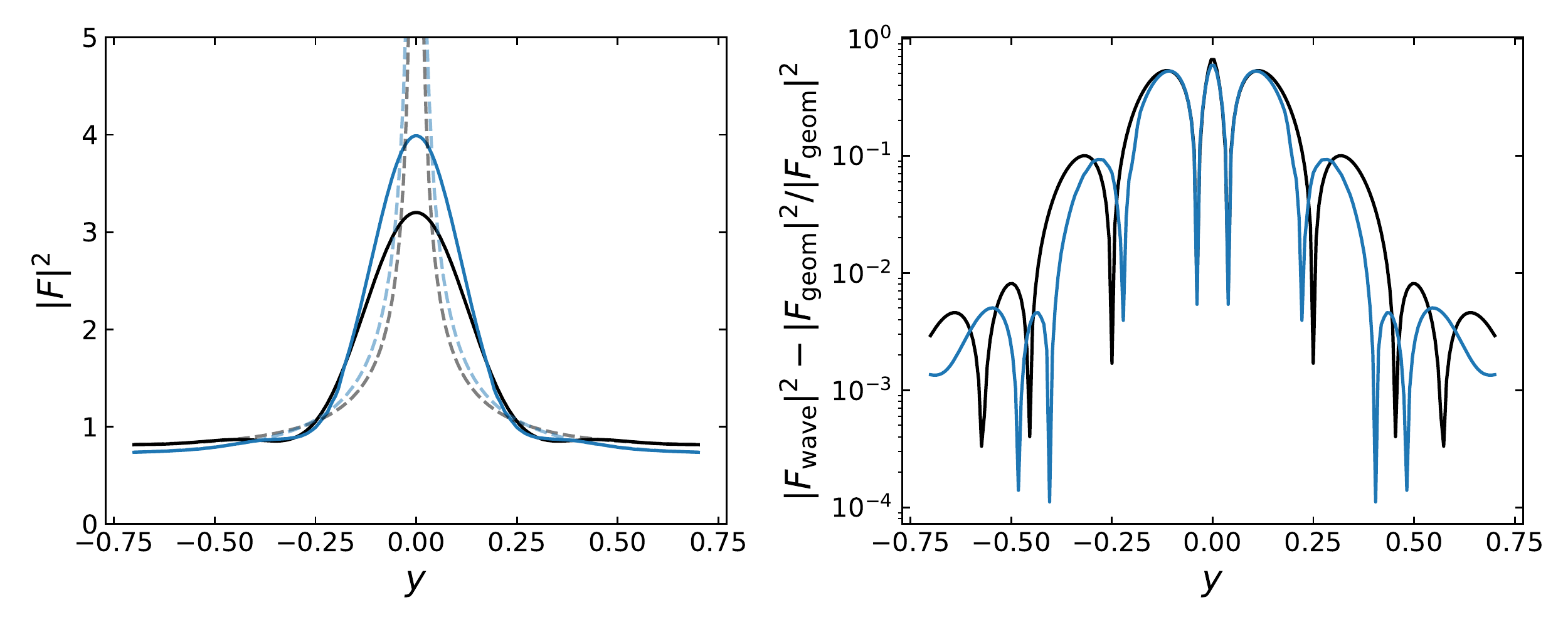}
    \caption{(Right) The wave optics magnification for an inverse quartic potential in blue, compared with the rational lens in black for $\alpha = 0.9$ and $\nu =10$. The geometric optics result is plotted in dashed lines. (Left) The fractional deviation of the wave optics magnification from the geometric magnification for the inverse quartic and rational lens.}
    \label{fig:quartic}
\end{figure}

Fig.~\ref{fig:quartic} shows the magnification for the rational and inverse quartic lens for $\alpha = 0.9$, as well as their deviation from geometric optics. Mathematically, the two lenses have Stokes lines at roughly the same location; however, the inverse quartic lens has two imaginary images that start to contribute. Despite the addition of an extra imaginary image, only one imaginary image dominates, yielding a very similar interference pattern as the rational lens. If one were to investigate even higher-order approximations to the Gaussian lens, one would find that despite a growing number of imaginary images that start to contribute at the Stokes line, the interference pattern is dominated by a single such image, yielding a qualitatively similar picture to the simple rational lens. Thus, while our focus has been on the rational lens, we argue that the effects we have been describing are generic.

We have also restricted our attention to the convergent rational lens ($\alpha > 0$). The divergent case may also be of interest, and, indeed, divergent lenses have been invoked to explain a variety of observed phenomena \citep[see e.g.][]{1998ApJ...496..253C, 2017ApJ...842...35C, 2018MNRAS.474.4637K, 2018MNRAS.481.2685D}. However, the divergent case has a more complicated caustic and Stokes line topology. Nevertheless, the general picture of imaginary images causing interference patterns in weak-lensing regimes is the same. A more detailed study of the divergent lens will be forthcoming in future work.

It is also important to note that we have taken the presence of regular oscillations in intensity (which we refer to as an interference pattern) to indicate the presence of multiple, interfering images. However, one may note that for any given intensity curve, it is always, in principle, possible to construct a lensing potential such that its geometric optics reproduces that intensity curve in a single-image regime. Thus, for an unknown lensing potential, regularly-spaced oscillations in intensity are not necessarily indicative of multiple images. Nevertheless, regularly-spaced oscillations are generic feature of interfering images, but require specifically tailored lensing potentials to be produced by a single image in geometric optics. Moreover, since the frequency dependence of geometric optics is fixed for any potential ($\psi \propto \omega^{-2}$ for plasma lenses), observations over multiple frequencies can be used to distinguish between geometric optics and the presence of genuine wave effects.

The question is, then, under what circumstances might wave effects in weak-lensing regimes be observable in data. Since the source must be an effectively coherent point-source to produce wave effects, many astrophysical sources are ruled out as candidates for this kind of observation. Sources that satisfy this coherency criterion for a wide range of potential applications include pulsars and fast radio bursts (FRBs).

Pulsars have been observed to undergo extreme scattering events (ESEs) caused by roughly AU-scale lenses, with electron column densities of order $\sim 0.01\,\mathrm{pc \, cm^{-3}}$ \citep[see e.g.][]{Coles2015, 2018MNRAS.474.4637K}. These events are consistent with isolated, highly asymmetric plasma lenses, that can be modelled with one-dimensional lensing potentials, such as the ones we have considered here. While the focus has been on modelling these events in strong-lensing regimes (indicated in data by the presence of spikes in intensity due to caustics), ESEs will also generically occur in weak-lensing regimes. Indeed, since the lens strength scales inversely with frequency squared, for any event observed in the strong-lensing regime, there will be a corresponding weak-lensing event in higher frequency bands. 

Using the rational lens model, we can compute the focal frequency, $f_\mathrm{focal}$, for typical physical parameters for the plasma lenses responsible for the ESEs presented in \citet{2018MNRAS.474.4637K}. The focal frequency is the frequency below which caustics form, and above which one is in the weak-lensing regime. That is, $f_\mathrm{focal}$ is the frequency above which $\alpha < 1$. From Eq.~\ref{eq:alpha}, we get
\begin{equation}
    f_\mathrm{focal} \sim 5.86\,\mathrm{GHz} \, \Big(\frac{a}{1\,{\rm AU}} \Big)^{-1} \, \Big(\frac{\overline{d}}{1\,{\rm kpc}} \Big)^{1/2} \, \Big(\frac{\Sigma_0}{0.01\,{\rm pc \, cm^{-3}}} \Big)^{1/2},
\end{equation}
where, again, $a$ is the size of the lens, $\overline{d} = D_d D_{ds} / D_s$, and $N_e$ is the peak electron column-density. Using Eq.~\ref{eq:nu}, we can convert this to the dimensionless frequency parameter, $\nu$, giving 
\begin{equation}
    \nu_\mathrm{focal} \sim 4500\,\Big(\frac{a}{1\,{\rm AU}} \Big) \, \Big(\frac{\Sigma_0}{0.01\,{\rm pc \, cm^{-3}}} \Big)^{1/2} \, \Big(\frac{\overline{d}}{1\,{\rm kpc}} \Big)^{-1/2}.
\end{equation}
When $\nu$ is large, the Eikonal limit becomes a better approximation of the underlying wave optics, and the contributions of the imaginary images are suppressed. However, as Fig.~\ref{fig:wavevgeom} shows, even $\nu \sim 1000$ can lead to a percent-level amplitude in the diffraction peaks. Nevertheless, smaller lenses, e.g. $a \sim 0.1\,{\rm AU}$, will lead to more easily observable interference effects in the weak-lensing regime. 

While many attempts to model ESEs have used divergent-lens models, convergent-lens models have also been proposed \citep{2012MNRAS.421L.132P}. However, as we noted above, Stokes lines in weak-lensing regimes are generic features of both convergent and divergent lenses, and their presence in ESEs will not be dependent on the precise details of the underlying lensing potential. 

For FRBs, the observational situation is a little more tenuous as ESEs have not been definitively observed for these sources. As \citet{2017ApJ...842...35C} argue, plasma lenses in FRB host galaxies with properties similar to the lenses responsible for pulsar ESEs in our galaxy are expected to lens FRBs and produce caustics. The lenses they describe will also generically lens FRBs in the weak lensing regime. However, since the time delays between images are large compared to the burst duration, the wave-optics formalism we have described does not apply. FRBs lensed by plasma lenses in our own galaxy, on the other hand, will have much smaller time delays. Since, in this case, $D_s \approx D_{ds} \ll D_d$, we have $\overline{d} \approx D_{d} \sim 1\,kpc$. Thus, the effects from imaginary images that we have been describing may also arise for FRBs being lensed by structures in the Milky Way. However, the path to detecting these effects is not as clear as in the case of pulsars. 

\section{Conclusion}
\label{sec:conclusion}

Stokes lines are generic features of a variety of lens models. At Stokes lines, imaginary images begin to contribute to the overall intensity modulation of coherent sources. This causes a discontinuity in the Eikonal approximation of wave optics. For very high frequency, this discontinuity is small, but as the frequency decreases the discontinuity grows larger, and the full diffraction integral must be calculated to give an accurate description of the wave optics.

In this paper, we have used a simple rational lens model as an example to study the effects of these imaginary images. We used Picard-Lefschetz theory in order to numerically evaluate the diffraction integral. This method has the advantage that, since each relevant image corresponds to a single Lefschetz thimble, of which the contribution to the total integral can be computed separately, one can compute the contribution of the relevant images individually in the full wave optics regime. In this way, Picard-Lefschetz theory can help bridge the gap between wave optics and the conceptual simplicity and power of geometric optics. 

Using the rational lens model, we have argued that imaginary images contain effectively as much information as the geometric images. It is well known that when multiple geometric images are present, interference effects allow for a direct measurement of the relative phases of the images, which, in turn, provides more information about the lens, helping to break degeneracies between the underlying physical parameters. However, we have shown that even in weak-lensing regimes, where only a single geometric image is present, interference patterns caused by interference with imaginary images can provide the same phase information as though multiple geometric images were interfering. We also discussed the possibility of observing these imaginary images in extreme scattering events of pulsars and FRBs.

\section*{Data Availability}
No new data were generated or analysed in support of this research.

\section*{Acknowledgements}
We thank Job Feldbrugge and Marten van Kerkwijk for useful  discussions. We receive support from the Ontario Research Fund - Research Excellence Program (ORF-RE), the Canadian Institute for Advanced Research (CIFAR), the Canadian Foundation for Innovation (CFI), the Simons Foundation, Thoth. Technology Inc, and Alexander von Humboldt Foundation. We acknowledge the support of the Natural Sciences and Engineering Research Council of Canada (NSERC), [funding reference number 523638-201,  RGPIN-2019-067, 523638-201]. Cette recherche a \'{e}t\'{e} financ\'{e}e par le Conseil de recherches
en sciences naturelles et en g\'{e}nie du Canada (CRSNG), [num\'{e}ro de
r\'{e}f\'{e}rence 523638-201, RGPIN-2019-067, 523638-201]. 




\bibliographystyle{mnras_sjf}
\bibliography{biblio} 



\appendix

\section{$n$-Dimensional Picard-Lefschetz Integrator}
\label{sec:plind}


This section describes the algorithm used to perform the Picard-Lefschetz integration, which was implemented in \texttt{Python}. In brief, the code takes in the gradient of the Morse function and an initial integration domain described by a simplicial surface. Then, by solving the steepest-descent flow equations (Eq.~\ref{eq:gradflow} with a negative sign) for the vertices of the simplices, the integration domain is deformed to lie along the relevant Lefschetz thimbles. Once the Lefschetz thimbles are found, the integration is carried out using the Grundmann-M{\"o}ller method \citep{doi:10.1137/0715019}. While in this paper we have only considered 1D oscillatory integrals, this general strategy works for arbitrarily high dimensions.

The code takes as input an integrand of the form $\exp{i S({\bm z})}$ where $S$ is analytic over the complex domain $\mathbb{C}^n$, and the complex gradient of the Morse function used to solve the flow equations. In order for the solution to the flow equations to be the Lefschetz thimbles of the intergand, the Morse function must be chosen to be $h({\bm z}) = \mathrm{Re} \{ i S({\bm z}) \}$, but the code allows these to be specified separately. The code also takes in an initial \texttt{contour} object, which is defined as a set of vertices and edges specifying an $n$-dimensional simplicial surface. This initial simplicial surface is the original integration domain prior to solving the flow equations, and will typically be chosen to span a subset of the real plane, $\mathbb{R}^n$. Other arguments include a number $\texttt{delta}$, which controls the maximum edge-length simplices are allowed to attain before triggering a mesh-refinement algorithm; $\texttt{thresh}$, which controls the minimum value of $h(z)$ vertices are allowed to achieve before being discarded; as well as \texttt{t\_init} and \texttt{tmax}, which are the initial timestep and maximum total time, respectively, for the particular ODE-solver implementation used to solve the flow equations.

The steps of the numerical algorithm is as follows:
\begin{enumerate}
    \item The initial simplicial surface, i.e. a mesh of connected simplices spanning the original integration domain, is specified along with the gradient of a Morse function, $\frac{dh}{dz^i}$. 
    \item The steepest-descent flow equations (Eq.~\ref{eq:gradflow} with a negative sign) are solved using the Runge-Kutta-Fehlberg 45 method \citep{fehlberg1969low} with an adaptive time-step, and initial time-step $\texttt{dt\_init}$. Note, Eq.~\ref{eq:gradflow} requires the specification of a metric; here we associate $\mathbb{C}^n$ with $\mathbb{R}^{2n}$ and use the standard Euclidean metric. 
    \item At each time-step, the edge lengths of the simplices, using the Euclidean metric, are computed. Any simplex with an edge length exceeding $\texttt{delta}$ are split in half along the offending edge. 
    \item At each time-step, the value of $h({\bm z})$ is computed for vertices of each simplex. Any simplex with a vertex, ${\bm z}_i$, such that $h({\bm z}_i) < \texttt{thresh}$ is removed. This step is necessary, since the downward flow lines of the Morse function will generically terminate at infinity or poles. In order to prevent all the points in the simplicial surface from flowing to infinity, the relevant domain is restricted to regions where $h({\bm z}) > \texttt{thresh}$.
    \item Once the ODE-solver has run for integration time \texttt{tmax}, the integrand $\exp iS({\bm z})$ is integrated over the final contour using a numerical implementation of the Grundmann-M{\"o}ller method from \texttt{quadpy} \citep{nico_schlomer_2021_4519699}. 
\end{enumerate}


\bsp	
\label{lastpage}
\end{document}